\documentstyle[epsf,aps,psfig]{revtex}
\renewcommand{\thefootnote}{\fnsymbol{footnote}}
\begin{document}
\newcommand{\be}{\begin{eqnarray}}
\newcommand{\dlq}{\lq\lq}
\newcommand{\ee}{\end{eqnarray}}
\newcommand{\ben}{\begin{eqnarray*}}
\newcommand{\een}{\end{eqnarray*}}
\newcommand{\beq}{\begin{equation}}
\newcommand{\eeq}{\end{equation}}
\renewcommand{\baselinestretch}{1.0}
\newcommand{\as}{\alpha_s}
\def\eq#1{{Eq.~(\ref{#1})}}

\def\ap#1#2#3{     {\it Ann. Phys. (NY) }{\bf #1} (19#2) #3}
\def\arnps#1#2#3{  {\it Ann. Rev. Nucl. Part. Sci. }{\bf #1} (19#2) #3}
\def\npb#1#2#3{    {\it Nucl. Phys. }{\bf B#1} (19#2) #3}
\def\plb#1#2#3{    {\it Phys. Lett. }{\bf B#1} (19#2) #3}
\def\prd#1#2#3{    {\it Phys. Rev. }{\bf D#1} (19#2) #3}
\def\prep#1#2#3{   {\it Phys. Rep. }{\bf #1} (19#2) #3}
\def\prl#1#2#3{    {\it Phys. Rev. Lett. }{\bf #1} (19#2) #3}   
\def\ptp#1#2#3{    {\it Prog. Theor. Phys. }{\bf #1} (19#2) #3}
\def\rmp#1#2#3{    {\it Rev. Mod. Phys. }{\bf #1} (19#2) #3}
\def\zpc#1#2#3{    {\it Z. Phys. }{\bf C#1} (19#2) #3}
\def\mpla#1#2#3{   {\it Mod. Phys. Lett. }{\bf A#1} (19#2) #3}
\def\nc#1#2#3{     {\it Nuovo Cim. }{\bf #1} (19#2) #3}
\def\yf#1#2#3{     {\it Yad. Fiz. }{\bf #1} (19#2) #3}
\def\sjnp#1#2#3{   {\it Sov. J. Nucl. Phys. }{\bf #1} (19#2) #3}
\def\jetp#1#2#3{   {\it Sov. Phys. }{JETP }{\bf #1} (19#2) #3}
\def\jetpl#1#2#3{  {\it JETP Lett. }{\bf #1} (19#2) #3}   
\def\epj#1#2#3{    {\it Eur. Phys. J. }{\bf C#1} (19#2) #3}
\def\ijmpa#1#2#3{  {\it Int. J. of Mod. Phys.}{\bf A#1} (19#2) #3}
\def\ppsjnp#1#2#3{ {\it (Sov. J. Nucl. Phys. }{\bf #1} (19#2) #3}
\def\ppjetp#1#2#3{ {\it (Sov. Phys. JETP }{\bf #1} (19#2) #3}
\def\ppjetpl#1#2#3{{\it (JETP Lett. }{\bf #1} (19#2) #3} 
\def\zetf#1#2#3{   {\it Zh. ETF }{\bf #1}(19#2) #3}
\def\cmp#1#2#3{    {\it Comm. Math. Phys. }{\bf #1} (19#2) #3}
\def\cpc#1#2#3{    {\it Comp. Phys. Commun. }{\bf #1} (19#2) #3}
\def\dis#1#2{      {\it Dissertation, }{\sf #1 } 19#2}
\def\dip#1#2#3{    {\it Diplomarbeit, }{\sf #1 #2} 19#3 }   
\def\ib#1#2#3{     {\it ibid. }{\bf #1} (19#2) #3}
\def\jpg#1#2#3{        {\it J. Phys}. {\bf G#1}#2#3}

\begin{flushright}

TAUP--2643--2000\\
\today
\end{flushright}
\vspace*{1cm} 
\setcounter{footnote}{1}
\begin{center}
{\Large\bf A manifestation of a gluon saturation in e-A DIS}
\\[1cm]
Eugene \, Levin $^{1,2}$ \,and\, Uri\, Maor  $^{1}$ \\

~

{\it $^1$ HEP Department, School of Physics and Astronomy } \\ 
{\it Tel Aviv University, Tel Aviv 69978, Israel } \\ 
{\it $^2$ Desy Theory, 22603 Hanburg, Germany}

\end{center}
\begin{abstract} 
This is a short presentation of our talks given at eRHIC Workshop at the
BNL. We give here a status report of our attempts to understand how 
gluon saturation will manifest itself in deep inelastic scattering with
nuclei. This summary reflects our current understanding and shows
directions 
of
our research  rather then a  final answer to the question.
Nevertheless, we are able to share with our  reader our tentative  answer
to the
question:``Why do we need to measure  DIS with nuclei and why these
data
will be complementary to the information obtained from  proton DIS".

\end{abstract}
\renewcommand{\thefootnote}{\arabic{footnote}}
\setcounter{footnote}{0}

\section{Introduction: what are the scales in photon-nucleus DIS?}
The main goal of these notes is to examine if and  how we can observe the 
 phenomenon of 
gluon saturation in DIS with nucleus. We present here only  a status
report of our attempts to clarify this subject, which is far away from 
being 
complete. It, rather, indicates the directions of our searches. Much 
 more work is needed  to  develop a reliable  approach so as 
to finalize our recommendations concerning experiments  the most sensitive
 to the
gluon saturation. 

We start with the general approach to photon-nucleus interaction,
developed by Gribov\cite{GRIB} who suggested following  time sequence  of
this process:
\begin{enumerate}
\item \quad First, the  $\gamma^* $  fluctuates into a hadron ( quark -
antiquark ) system well before the interaction with the target;
\item \quad Then the converted quark-antiquark pair ( or hadron system)
interacts with the target.
\end{enumerate} 
Generally,  these two stages result in the following formula for
the cross section
\beq \label{GF}
\sigma_{tot}( \gamma^*  + A )\,\,=\,\,\sum_n
\,\,|\Psi_n|^2\,\,\sigma_{tot}(n + A; x)\,\,,
\eeq
where $\Psi_n$ is the wave function of the system, produced in the first
stage of the process.
\subsection{ Separation scale $\mathbf{r^{sep}_{\perp} \,\approx\,1/M_0}$}
This scale is a typical distance which separates the pQCD approach  from
the non-perturbative one. Roughly speaking, for shorter distances than
$r^{sep}_{\perp}$,  the QCD running coupling constant can be considered as
a small parameter while for longer distances $\alpha_S(r_{\perp})$ is
large
and we cannot use the powerful methods of pQCD.   Table 1 demonstrates how
this scale works in
our particular model to incorporate the long distance physics
\cite{GLMG,AGL,OSAKA}.
\newpage
\begin{table}
\centerline{\bf Table 1}
\begin{tabular}{ l l l}
 {\large Perturbative QCD} & $\longrightarrow$ &
{\large  non-perturbative QCD}\\
 & & \\
{\large short distances} &  $\longrightarrow$ &  {\large long distances} 
\\
 & & \\
{ \Large $ r_{\perp} \,\,<\,\,$} & {
 \Large $ r^{sep}_{\perp}$}
&\,\,\,\,\,\,\,\,\,\,\,\,\,\,\,\,{  \Large
$<\,\,
r_{\perp}
$}\\
   & & \\
DOF: colour dipoles \cite{MU94}& $\bullet$ & DOF: constituent
quarks\cite{LF} \\
 &  & \\
 $\Psi_n$:  QED for virtual photon  & $\bullet$ & $\Psi_n$: generalized
VDM for
hadronic system\\ 
 & &\\
$\sigma_{tot}(n,x) = \sigma (r^2_t,x)$ & $\bullet$ &
$\sigma_{tot}(n,x) =
\sigma(q q \rightarrow q q; x)$ \\
 & & \\
Glauber- Mueller  Eikonal \cite{MU90}  for $\sigma (r^2_{\perp},x)$ &
$\bullet$ &
Regge phenomenology
for $\sigma(q + q \rightarrow q + q; x)$\\
\end{tabular}  
\end{table}

It is important to notice that the separation scale mostly relates to the 
produced hadronic ($q \bar q $) ) system and  
does not depend on
the properties of the target ( in particular,  the  atomic number ).
From Table 1 one can write for short distances ( $r_{\perp}\, < \,
r^{sep}_{\perp}$ ) 
\beq  \label{SDXS}
\sigma_{tot} ( \gamma^* p ) =  \int d^2 r_{\perp}  
\int^1_0 \,d z\, | \Psi( Q^2; r_{\perp},z ) |^2 \,\sigma_{tot} (
r^2_{\perp}, x
)\,.
\eeq

\subsection{ Saturation scale $\mathbf{r^{sat}_{\perp}
\,\approx\,1/Q_s(x;A)}$}
   
At low $x$ and at $r_{\perp} <  r^{sep}_{\perp}$ we believe \cite{SAT} 
that
{\it The system of partons always
passes the stage of hdQCD
( at shorter distances ) before it goes to the black box, which we call
non-perturbative QCD,  and which, in  practice, we describe in old fashion
Reggeon phenomenology.} At the  hdQCD stage we have to observe a parton
system
with sufficiently small typical distances ( $
r^{sat}_{\perp}\approx\,1/Q_s(x;A) $ ) at which the QCD coupling constant
is still small ($\alpha_S(r^{sat}_{\perp}) \ll 1 $),  but the density of
partons is so large that we cannot use
the pQCD methods in our calculations.  The picture of the parton
distribution in the transverse plane is shown in Fig. {~\ref{pcsd}}.

The estimate of the value for the saturation scale is obtained
\cite{SAT} from the equation
\beq \label{SAT}
\kappa\,\,\,=\,\,\,\frac{3\,\pi^2 \alpha_S \,A}{2
Q^2_s(x)}\,\times\,
\frac{xG(x,Q^2_s(x))}{\pi\, R^2_A}\,\,=\,\,1\,\,,
\eeq
where $A$ and $R_A$ are the atomic number and radius of the nucleus.
\eq{SAT} has a simple physical meaning giving the probability of the 
interaction between two partons in the parton cascade. Namely, such an
interaction will stop the increase of the parton density due to parton
emission, which is  included in the DGLAP evolution equations
\cite{DGLAP}.

It is important to notice that the saturation scale strongly depends on
$A$ ($Q_s(x;A) \propto A^{\frac{1}{6}}\cite{MULA}  \div
A^{\frac{1}{3}}\cite{KOLE} $ ).
\subsection{The theory status}
In eA deep inelastic scattering we want to find the high  density
parton system which is a non-perturbative  system but which can be treated
theoretically. It should be stressed that the  theory of hdQCD is in a
very
good shape now. Two approaches have been developed for hdQCD:  the
first
one\cite{PTHEORY} is based on pQCD ( see GLR and Mueller and Qiu in Ref.
\cite{SAT})  and
on the dipole degrees of  freedom \cite{MU94}, while  the
second\cite{ELTHEORY}  uses the
effective Lagrangian, suggested by McLerran and Venugopalan \cite{SAT}.
As a result of this intensive work we know now the nonlinear equation
which governs the QCD evolution in the hdQCD region \cite{EQ}.
We have not developed simple methods  to estimate an effect of  hdQCD on the
experimental observables and have to use a model approximation, but we
want to emphasize that this is a
temporary stage of our theory which will be overcome soon.
\section{HERA: results and puzzles.}
We start answering the question:``why do we need a nuclear target to find
the hdQCD phase" with a summary of  what we have learned from HERA. 

\begin{figure}
\begin{center}
\epsfxsize=9cm
\leavevmode
\hbox{ \epsffile{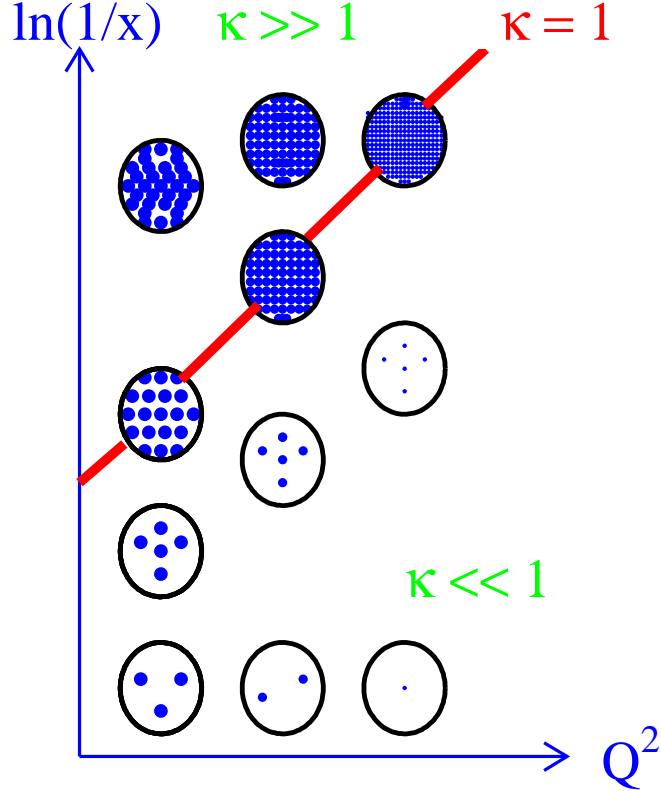}}
\end{center}
\caption{The parton distribution in the transverse plane. The curve shows
the saturation scale $Q_s(x;A)$}
\label{pcsd}
\end{figure}

\begin{itemize}

\item \quad $\mathbf{F_2}$ - the most striking and significant result from
HERA is
the increase of $F_2$ at low $x$ \cite{HERADATA}. The interpretation of
the $F_2$ data in terms of the DGLAP evolution equations leads to
sufficient
large value and  a steep behaviour of the gluon structure function at low
$x$.  $xG(x,Q^2)$ turns out to be so large that $\kappa$, our   new order
parameter,  exceeds unity in a significant part of the  accessible phase
space (
see
Fig. \ref{kappa}).

\begin{figure}
\begin{center}
\epsfxsize=7cm
\leavevmode
\hbox{ \epsffile{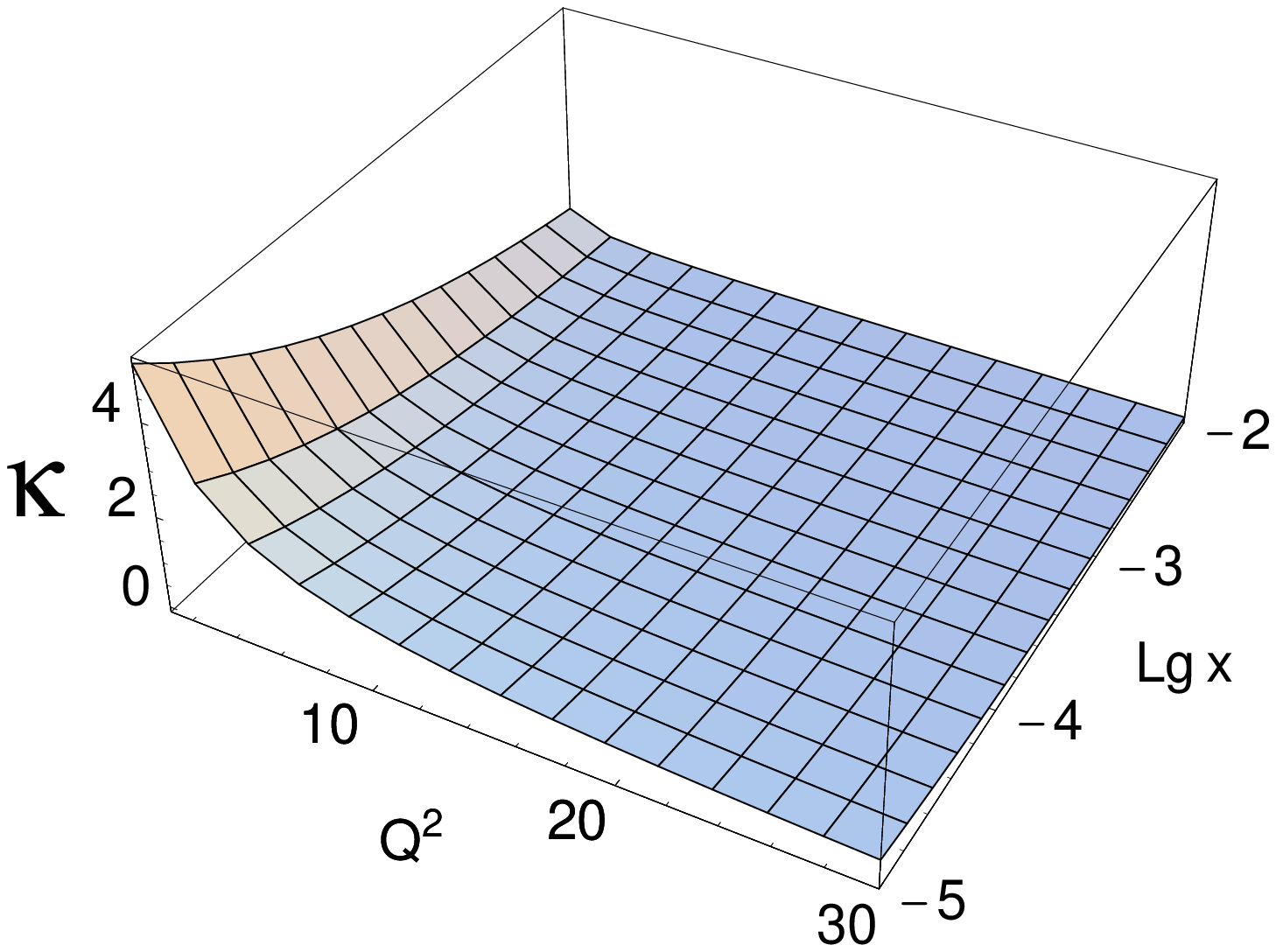}}
\end{center}
\caption{}
\label{kappa}
\end{figure}

\item \quad {\bf  Diffractive production} - three important results have
been observed at HERA: (i) the diffractive production gives a substantial
part of the total cross section, about 10 $\div$ 15 \% at $Q^2 \approx 10
\,GeV^2$; (ii) the energy behaviour of the diffractive cross section has
an intercept larger that the intercept of the soft Pomeron, namely,
$\sigma_{diff} \propto (1/x)^{2 \Delta_P}$ with $\Delta_P >
\Delta_{softP}\,\approx 0.1 $ \cite{DL}, and (iii) the ratio
$\sigma_{diff}/\sigma_{tot}$ is a constant versus energy in HERA kinematic
 region. From Fig. \ref{pcsd}, one can see that a hadron looks as a
diffractive grid with typical size of the order of $r^{sat}$. Therefore,
we expect that diffractive processes  originate from a rather small
distances. This fact leads to a natural explanation of the energy
behaviour of the diffractive cross section. 

\item \quad {\bf Matching between soft and hard processes.} The
experimental data on $\gamma^* p $ cross section at  small $Q^2$
allows to test different models for the matching of the soft and hard
interactions. 

\item \quad The dedicated  beautiful {\bf  measurement of the $F_2$
slope}
($d
F_2/\ln Q^2$) gives us a hope to find the saturation scale by observing
the movement of the maxima in $Q^2$ - behaviour at fixed $x$. The
experimental data show a considerable deviation fron the DGLAP analysis at 
$Q^2 \leq 1 \div 3 \,GeV^2$. 
However,
the current data can be described in  two different ways, either due to a
gluon saturation or due to a probable matching between soft and hard at
rather large
momenta ( about 1 - 2 GeV ) \cite{OSAKA}. It should be noticed, however,
that the J/$\Psi$ production can be easily described taking into account
shadowing corrections confirming a gluon saturation hypothesis
\cite{OSAKA}.

\end{itemize}

\begin{figure}
\begin{center}
\epsfysize=8cm
\leavevmode
\hbox{ \epsffile{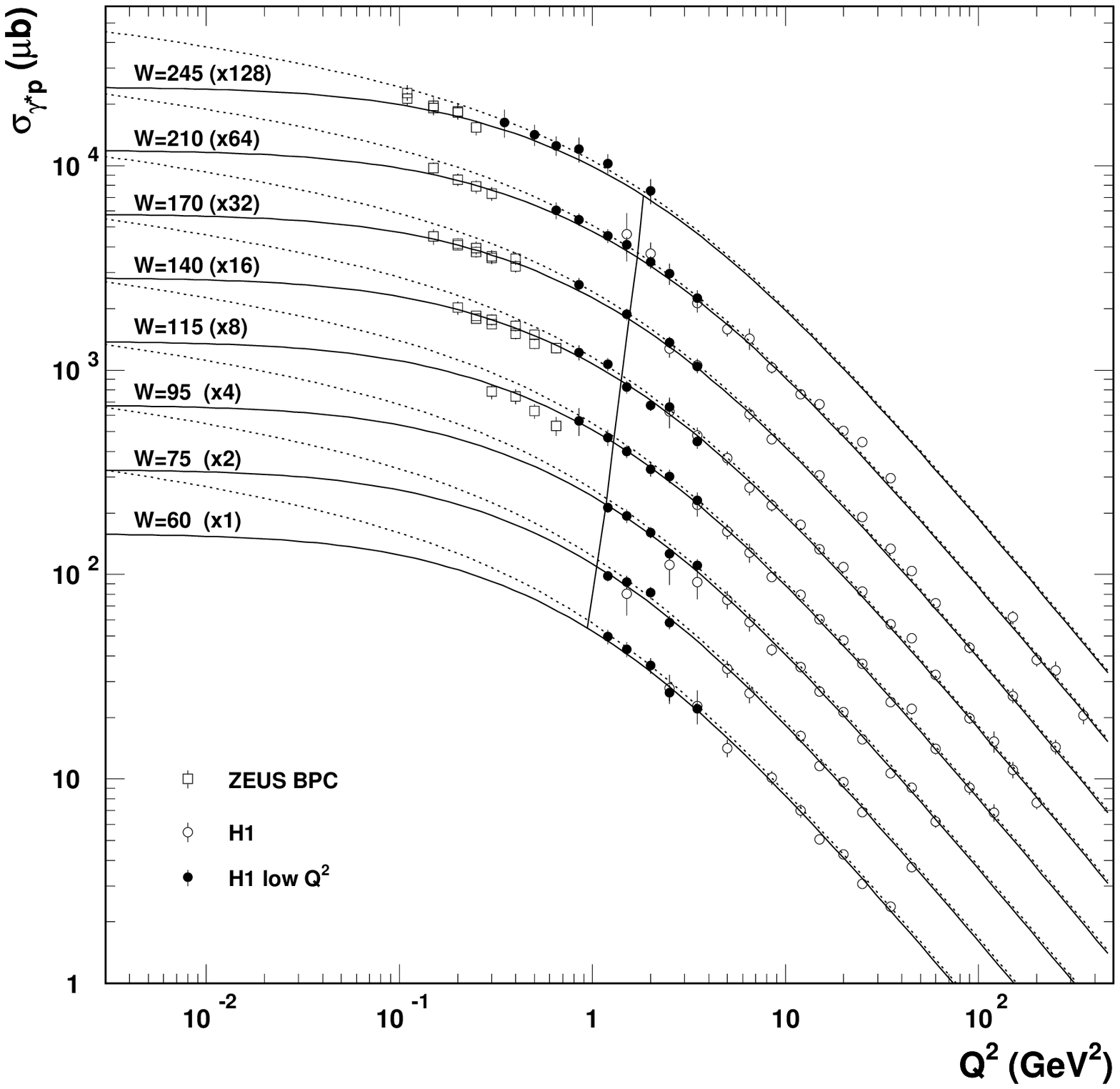}}
\end{center}
\caption{}
\label{gw1}
\end{figure}

\begin{figure}
\begin{center}
\epsfxsize=7cm
\epsfysize=6cm
\leavevmode
\hbox{ \epsffile{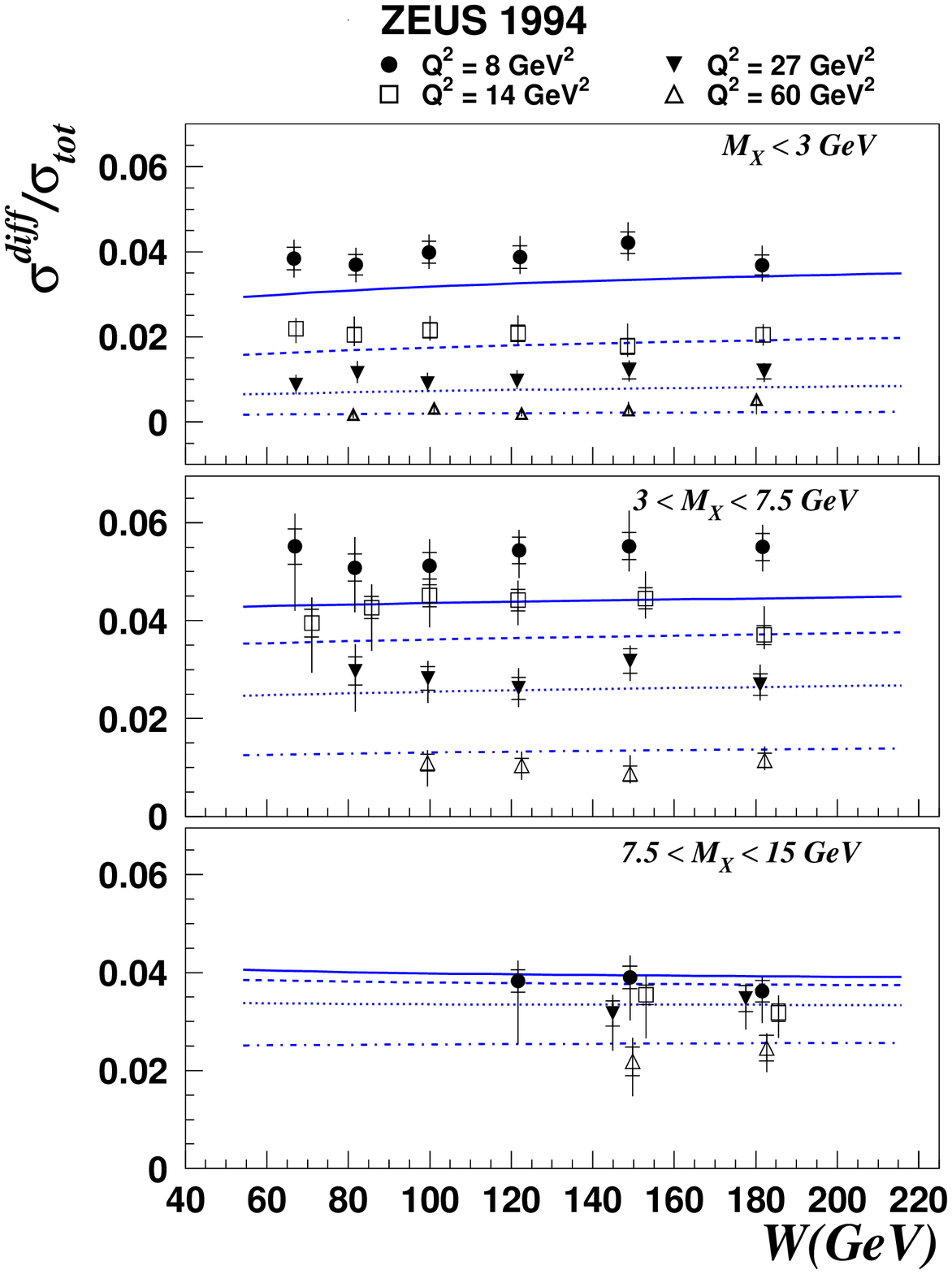}}                               
\end{center} 
\caption{}                                                            
\label{gw2}                       
\end{figure}

We listed above the most important HERA observations  which indicate a
possible
saturation effect. To illustrate this fact we will demonstrate that  a
simple parameterization of  
Golec-Biernat and Wusthoff\cite{GW}, which includes the saturation,
works  well. They  found an elegant
phenomenological model for $\sigma(r^2_{\perp},x)$ in \eq{SDXS} which is
able to describe all experimental data using only three parameters
\cite{GW}. In this model 
\beq \label{GW}
\sigma_{dipole} (r_{\perp},x)\,\,\, =\,\,\,\sigma_0 \,\left(\,1 - exp^{-
\frac{r^2_{\perp}\,Q^2_0}{(x/x_0)^{\lambda}}}\,\right)\,\,,
\eeq
with $\sigma_0 = 23.03 mb$,  $Q^2_0 = 1 GeV^2$, $ x_0 = 0.0003$  and
$\lambda = 0.288$. Figs. \ref{gw1} and \ref{gw2} show the quality of this
description. 
Note that even though the above description is impressive, it
cannot fix the value of the saturation scale from the data which is too
 constrained by the kinematics ( see Ref. \cite{OSAKA}
for details).

Thefore, we can conclude that the saturation hypothesis is compatible 
with all experimental data. However, the puzzling situation is that 
the same data can  be described fron a different point of view without a 
saturation scale in  the
standart DGLAP evolution equation for $Q^2 > 1 GeV^2$ and the soft
phenomenology for $Q^2 < 1 GeV^2$.  We  do not claim that
it  is  a reasonable or smooth parameterization of the data but
Donachie-Landshoff 
mixture of soft and hard Pomeron  shows that we  can produce  such a
model.

 Thus in order to fix the saturation scale and to descriminate between
competing models, we need either to reach a much  smaller values of $x$ (
higher energy) or to use a new target.  Realistically, we can conclude
that

\begin{minipage}{16cm}{\it We need DIS with nuclei to check whether  the
indications on saturation effect at HERA are really true.}
\end{minipage}

\section{Scale of the saturation effect for DIS with nuclei}
\subsection{Asymptotic predictions.}
Let us start with listing the asymptotic predictions of our approach
\cite{GLMG,AGL,OSAKA} which is based on the Glauber-Mueller formula for
$\sigma(r^2_{\perp},x)$ in \eq{SDXS} \cite{MU90}.

\begin{itemize}
\item \quad At fixed $r_{\perp} $ and at $x \rightarrow 0$
$$  \sigma^{dipole}_{tot} \,\, 
\longrightarrow\,\,2\,\pi\,\,\left( R_A
\,\,+\,\,\frac{h}{2}\,\ln(Q^2_s(x;A)/Q^2)\,\right) $$
where $R_A$ is the nucleus radius and $h$ is the surface thickness in the
Wood-Saxon nucleon density\,\,;
\item \quad In the same limit $ \frac{d F^A_2}{d \ln Q^2} \,\,
\longrightarrow\,\,F^A_2
\,\left(\,1 \,\,-\,\,\frac{h}{R_A}\,\right)\,\,
\propto\,\,\,Q^2\,\,R^2_A$;
\item \quad  The ratio of the diffraction to the total cross sections
should depend on energy only weakly \cite{MK};
$$
\frac{\sigma^{diffraction}_{tot}( \gamma^* A)}{\sigma_{tot}(\gamma^* A )}
\,\,\approx\,\,\,Const( W ) \,\, \longrightarrow \,\, (
slowly )\,\,\,\,\frac{1}{2}; $$

\item \quad  The energy behaviour of $\sigma^{diffraction}(
\gamma^* A )$ is determined by short distances $ r_{\perp}
\,\approx\,1/Q_s(x;A)$;

\item \quad  The high density effects should be stronger in the
diffractive channels.    

\end{itemize}

\subsection{$\mathbf{xG_A(x,Q^2)}$}
In Fig.5 we present our calculation of the gluon structure function for
different nuclei. Fig.5-d gives a glimpse at what we are taking into
account in our approach. Figs. 5-a - 5-c show the comparison of our
calculations, based on  the Glauber-Mueller formula, with the solution of
the full equation for hdQCD \cite{EQ} \footnote{Actually, the equation,
suggested in Ref. \cite{AGL} was solved and plotted in Fig.5 as the
asymptotic solution, but this equation in the double log approximation
coincides with the correct one \cite{EQ}.}

\begin{figure}[hptb]
\begin{center}
\begin{tabular}{ c c}
\psfig{file=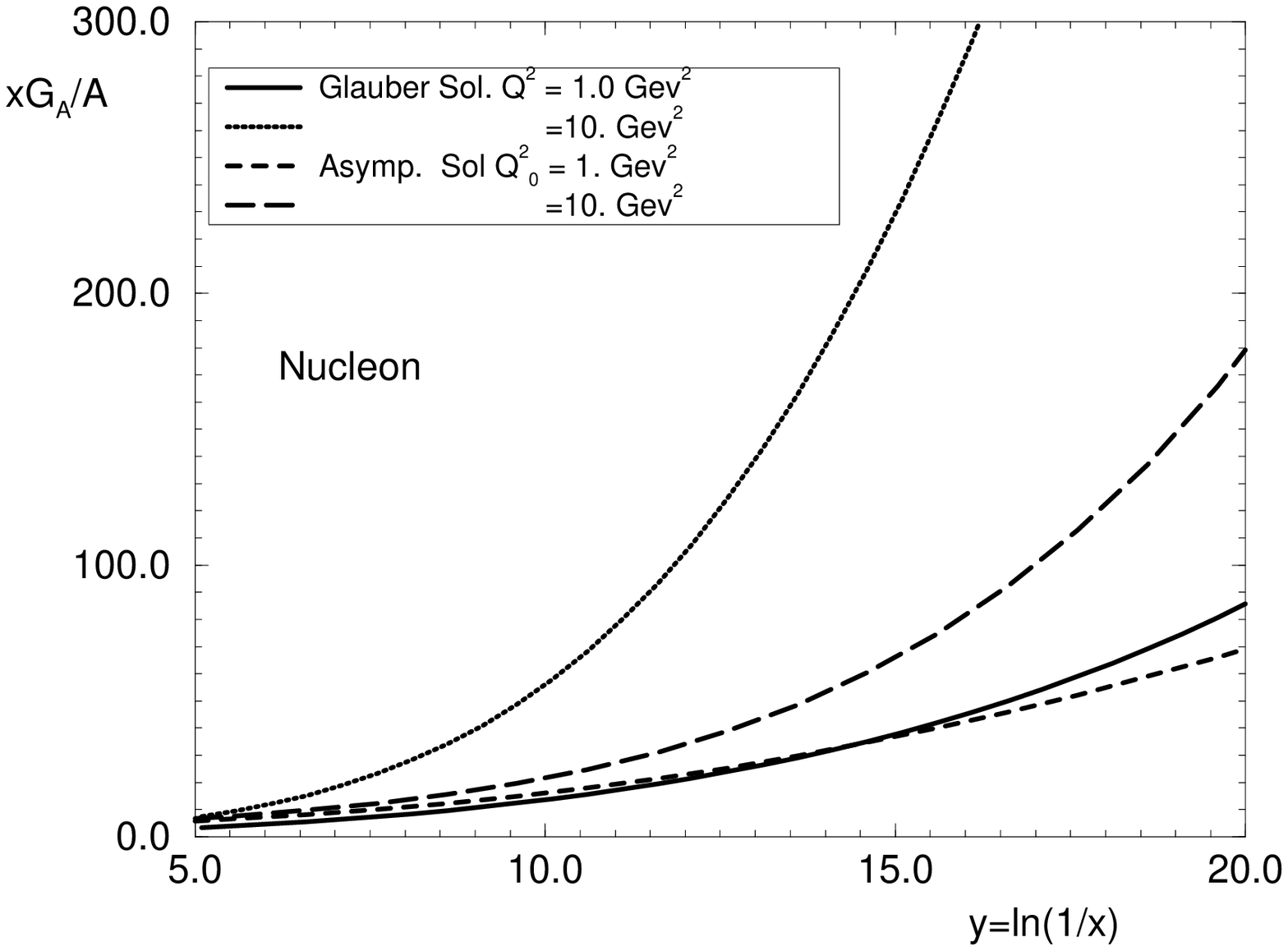,width=70mm,height=65mm} &
\psfig{file=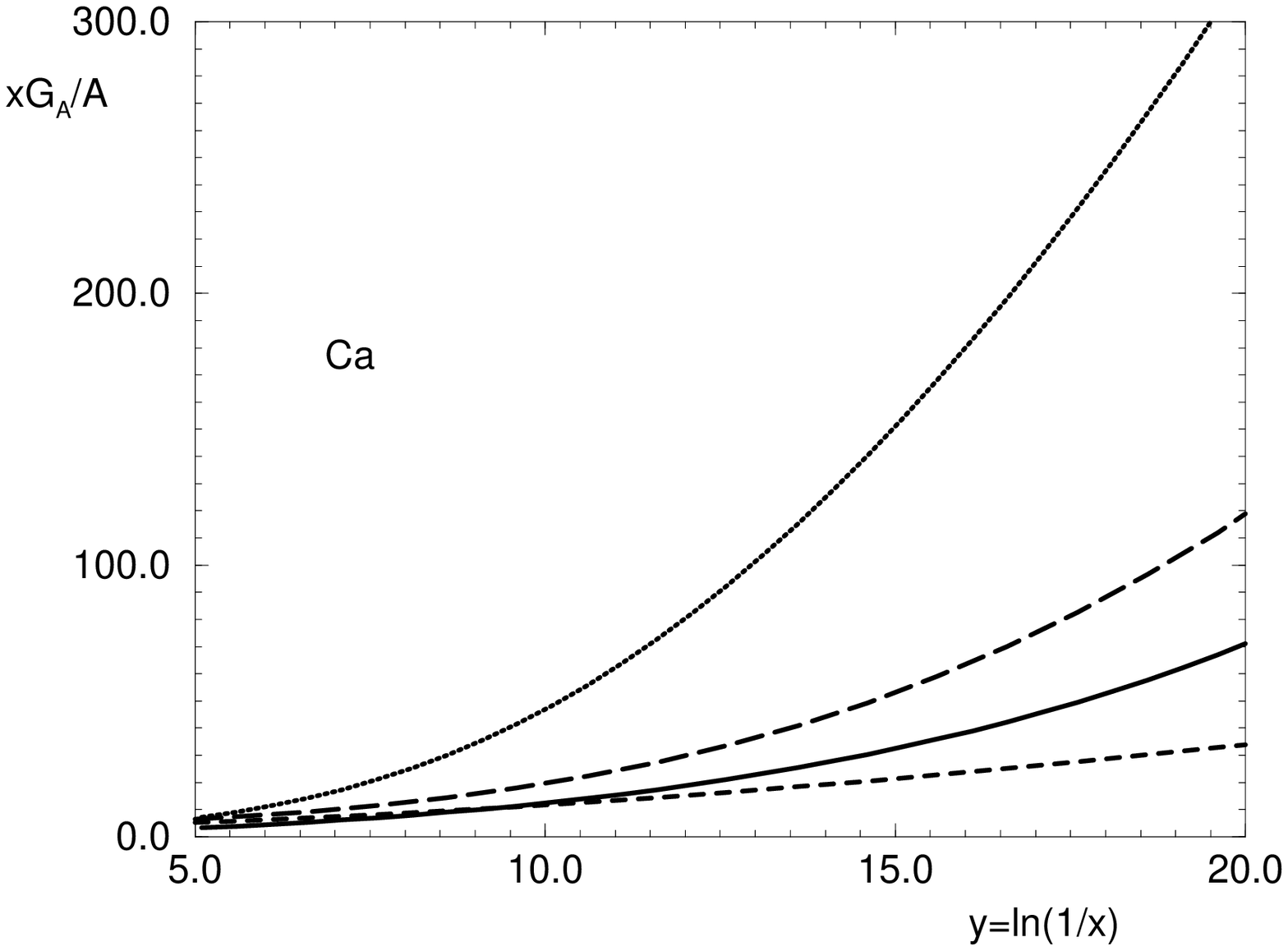,width=80mm,height=65mm}\\
 Fig.5-a & Fig.5-b  \\
\psfig{file=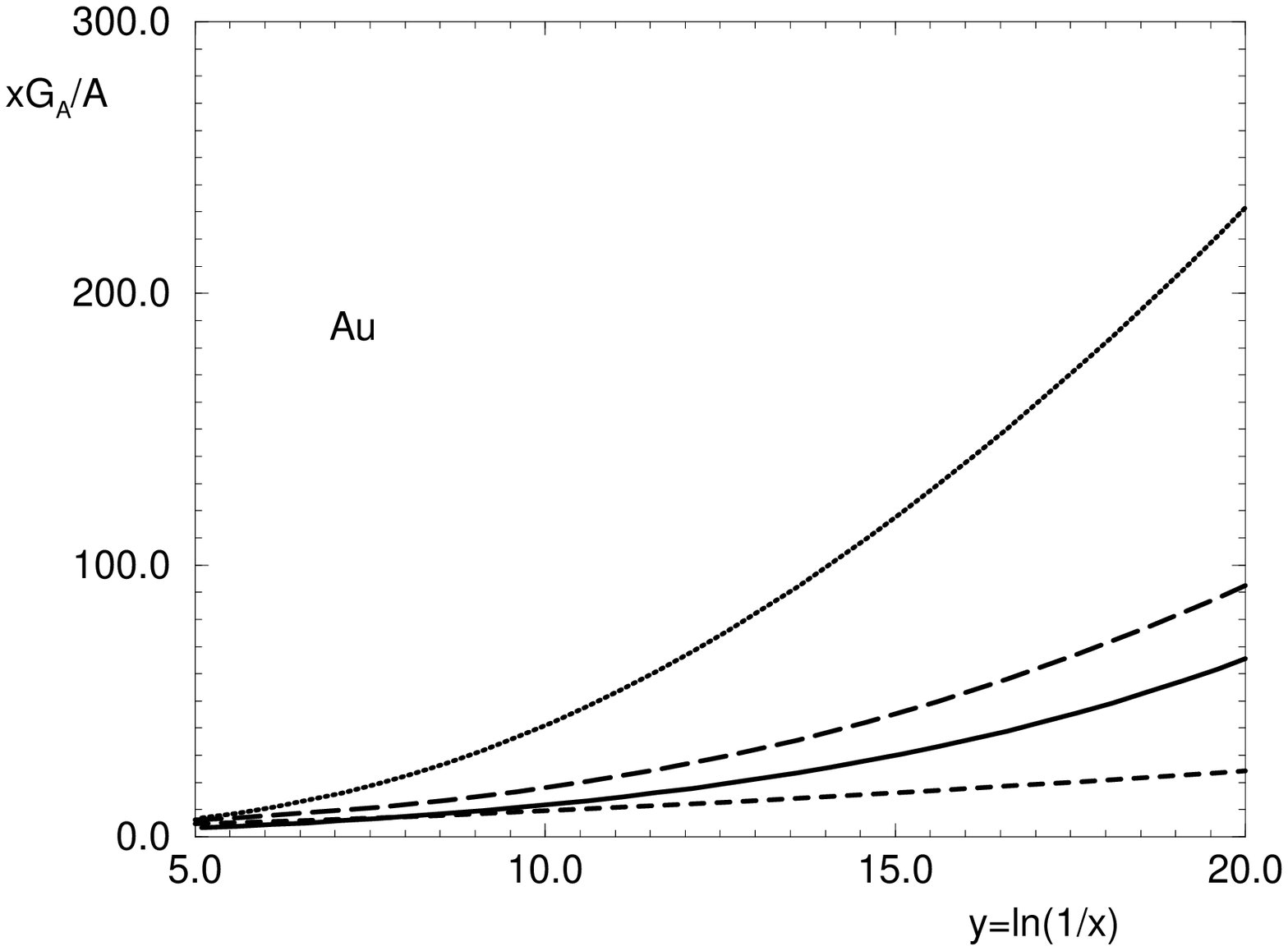,width=80mm,height=65mm}  
&\psfig{file=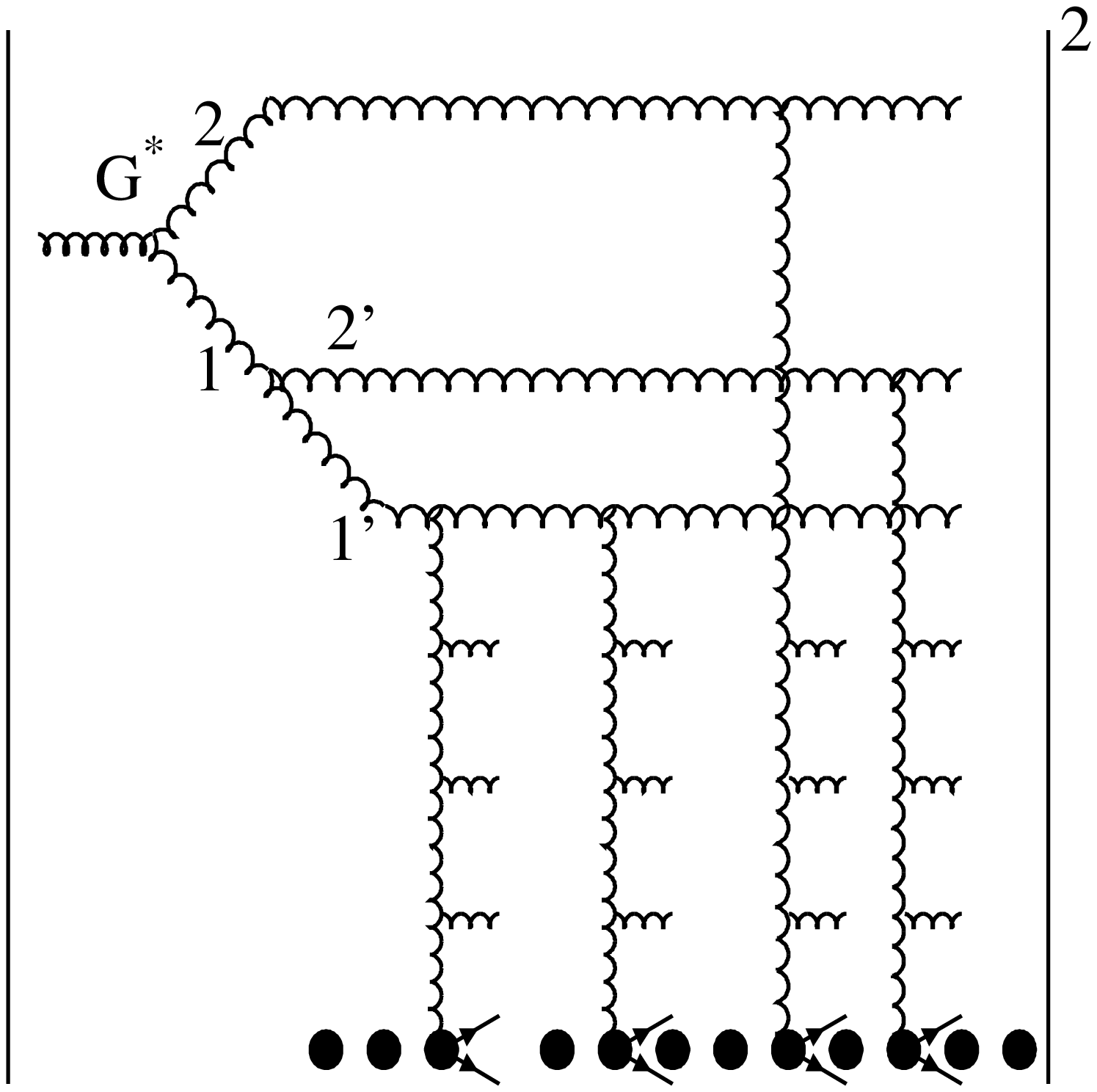,width=80mm,height=65mm}\\
 Fig. 5-c & Fig.5-d\\
\end{tabular}
\end{center} 
\caption{}
\label{nxg}
\end{figure}

We can derive two conclusions from Fig.5: (i) the saturation effect is
much stronger for a nucler target than for a nucleon, and (ii) our model
underestimates the value of the effect for $Q^2 \approx 1 GeV^2$. 
Unfortunately, we have not finished our estimates for $F_2$ for DIS with
nuclei.

\subsection{R\,=\,$\mathbf{\frac{\sigma^{diffraction}_{tot}}{\sigma_{tot}}}$}

This ratio shows us how we are  close, or how we are far away, from the
asymptotic regime since at very high energy it should be equal to
$\frac{1}{2}$.  In Fig.6 we plotted our calculations for this ratio
\cite{DDOUR}. One can see that the ratio is larger than for the proton
target ( see Fig. \ref{gw2} ) but it is still smaller than the limiting
value of $\frac{1}{2}$. This is a very encouraging fact for experiment
since
we do not want to measure a black disc limit which is not sensitive to the
theoretical approach. In other words, any model or any theoretical
approach will give  the  unitarity limit which we call `` black disck
limit".
A ll our theoretical QCD prediction are related to the form of
the transition from  pQCD to the  ``black disc limit".

\begin{figure}
\begin{tabular}{cc}
$\mathbf{A}${\bf =30}
&
$\mathbf{A}${\bf =100}\\
\psfig{file=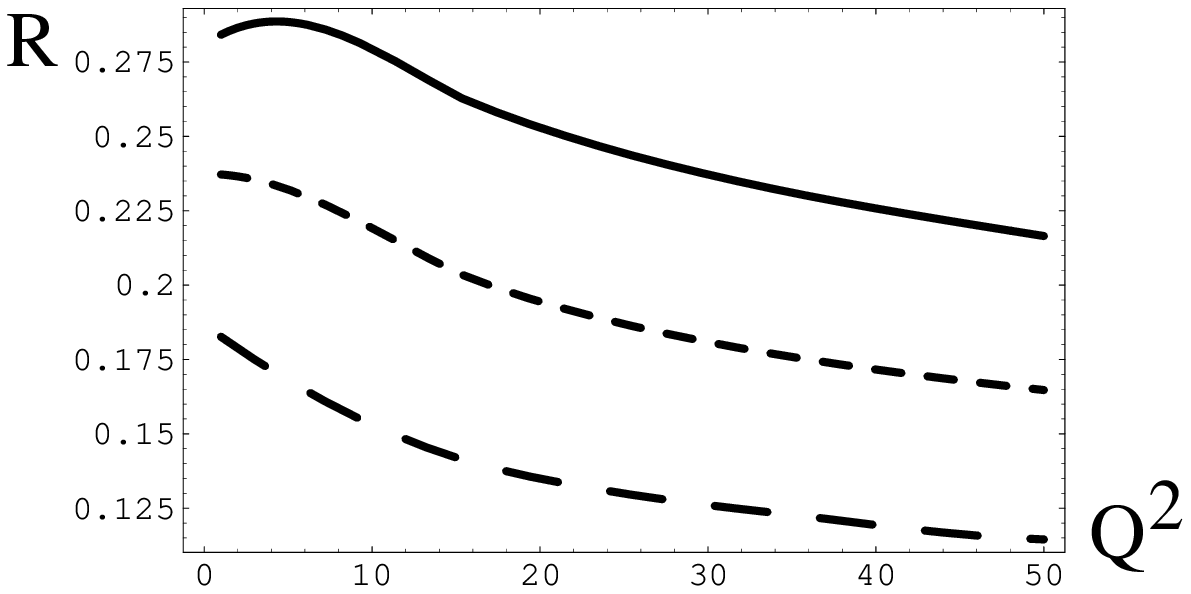,width=70mm,height=40mm}
&
\psfig{file=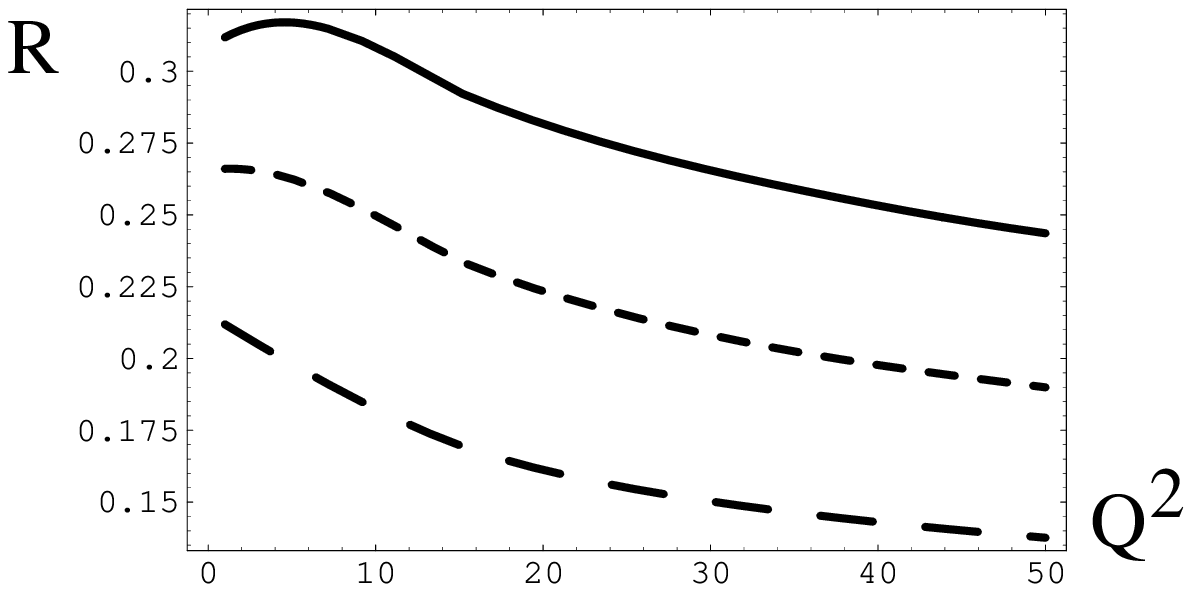,width=70mm,height=40mm}\\
$\mathbf{A}${\bf =200}
&
$\mathbf{A}${\bf =300}\\
\psfig{file=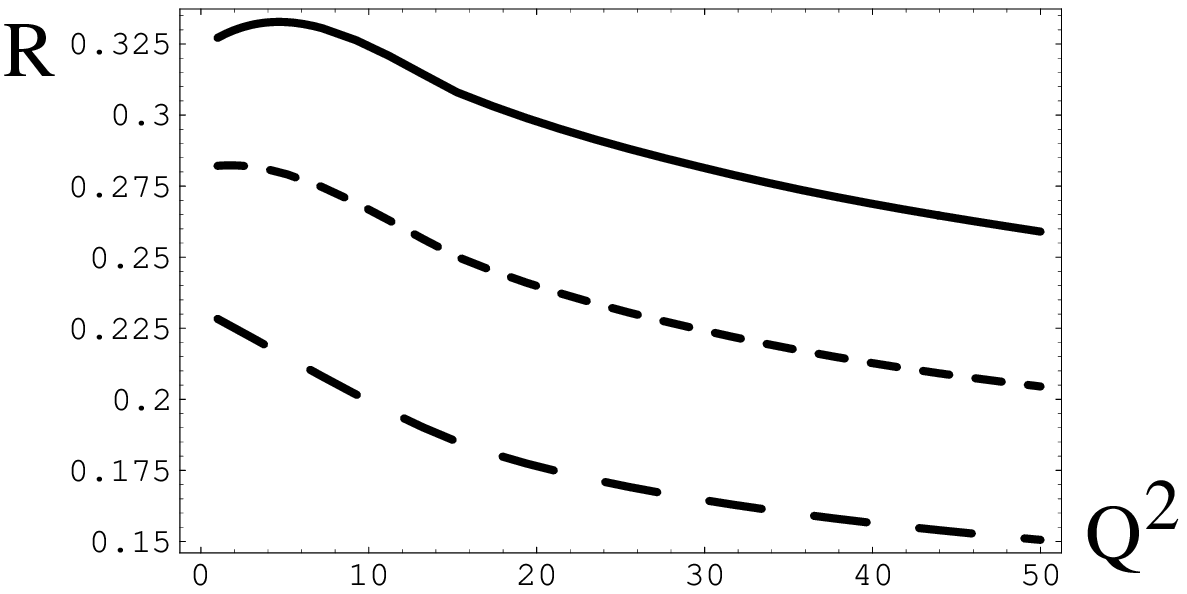,width=70mm,height=40mm}
&
\psfig{file=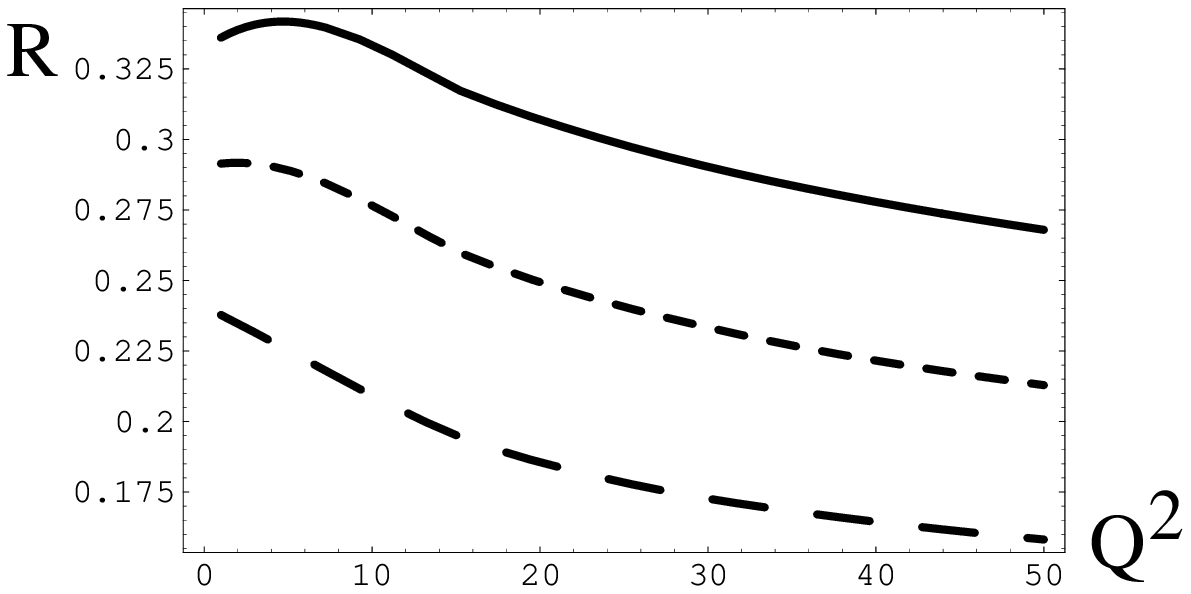,width=70mm,height=40mm}
\label{fig1}
\end{tabular}
\caption{}
\end{figure}
\subsection{ $\mathbf{F_2}$ slope (
$ \mathbf{\frac{\partial F^A_2(x,Q^2}{\partial \ln\,Q^2}}$) }
\begin{figure}
\begin{center}
\epsfxsize=10cm
\leavevmode
\hbox{ \epsffile{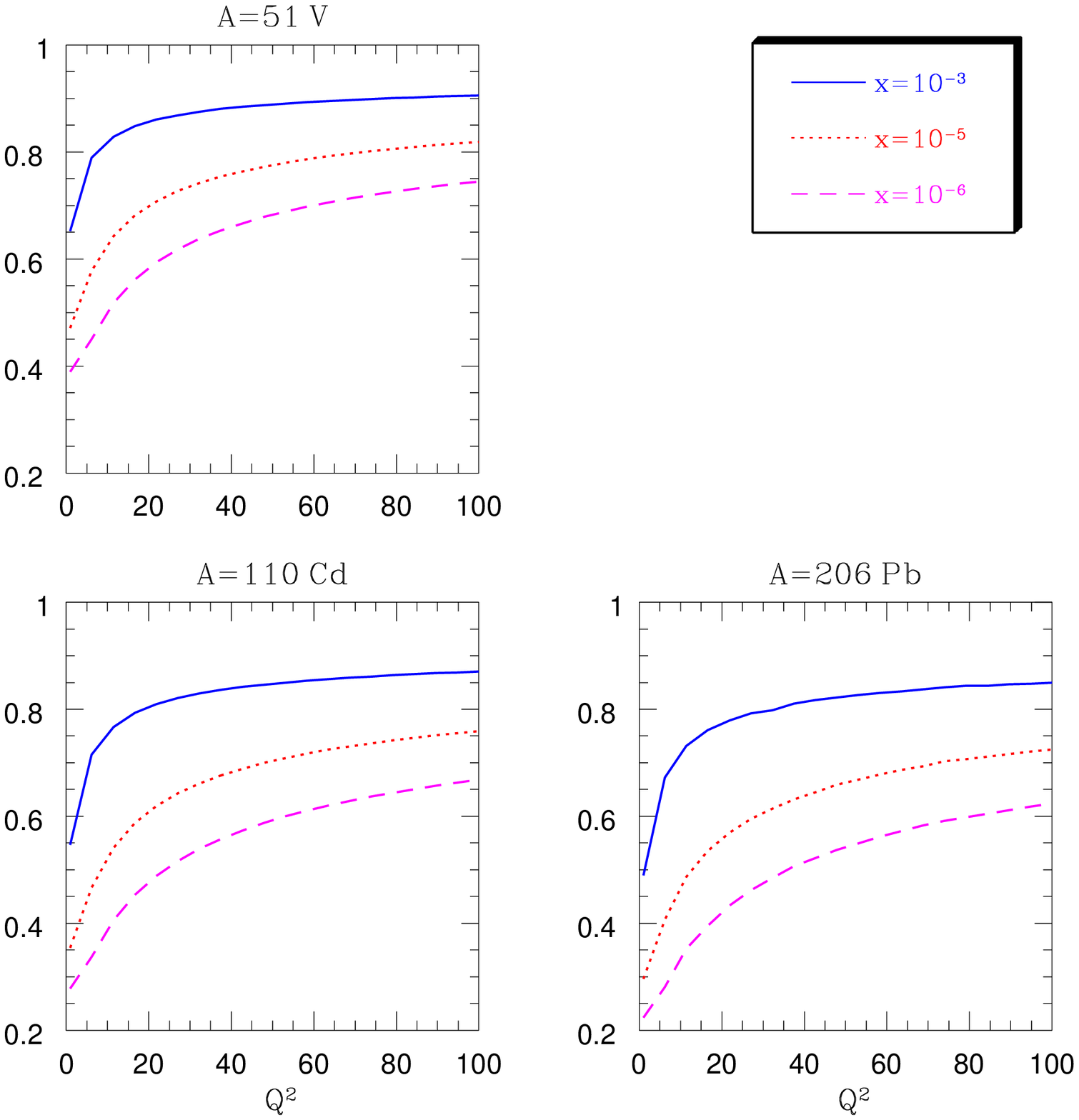}}
\end{center}
\caption{}
\label{slp1}
\end{figure}

\begin{figure}
\begin{center}
\epsfxsize=10cm
\leavevmode
\hbox{ \epsffile{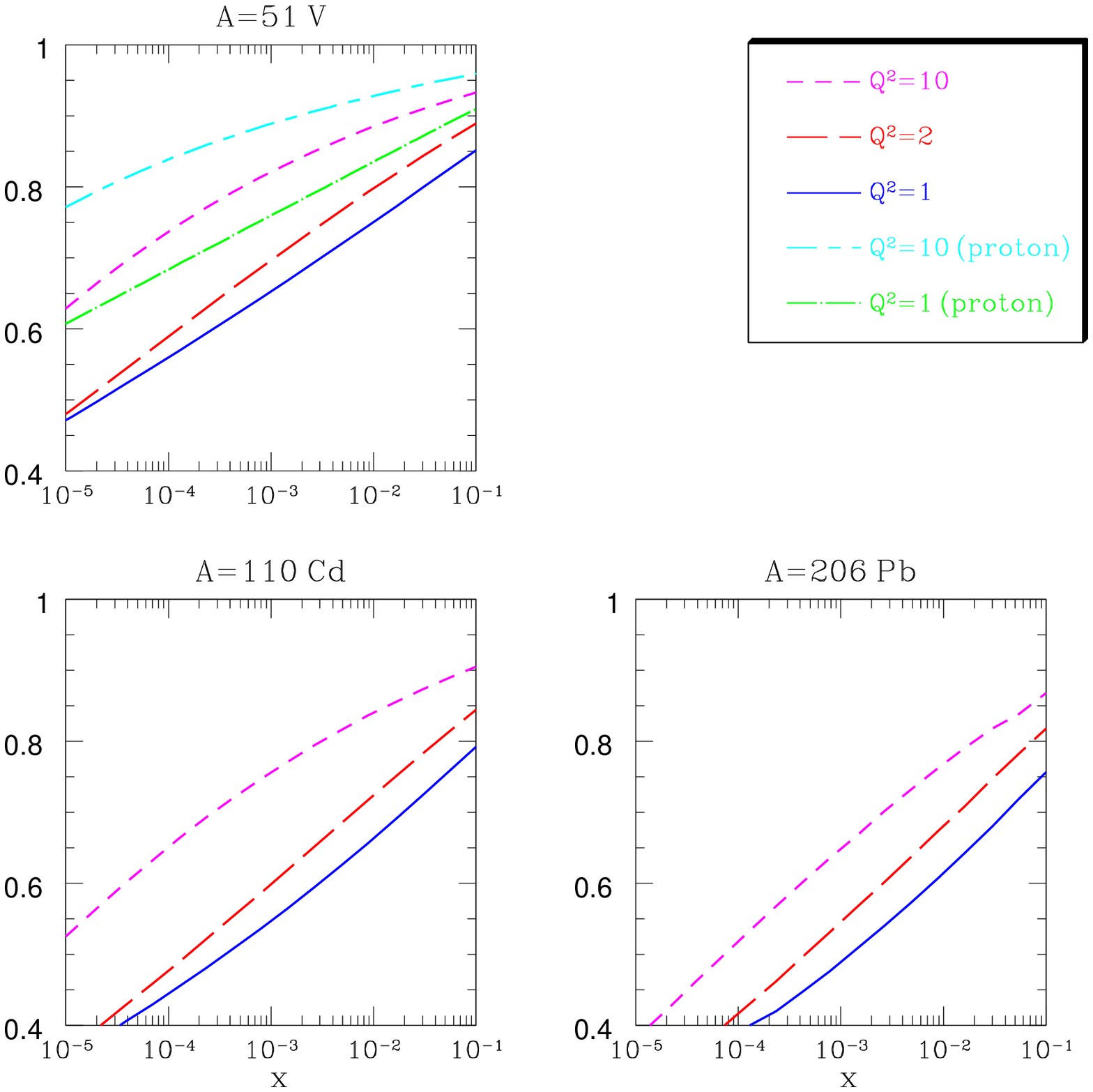}}    
\end{center}
\caption{}                                                            
\label{slp2}                        
\end{figure}

As we have mentioned,  gluon satruration leads to a maximum in the
$Q^2$ dependence of 
$F_2$ slope at  $Q^2= Q^2_s(x;A)$ fora  fixed value of $x$. Such a 
maximum has not been seen in the HERA data and has not been anticipated in 
our estimates of the slope. However, the numerical value of the
gluon saturation is rather  for a nucleus target.
Figs. \ref{slp1} and \ref{slp2} display  the possible experimental
effect. In these figures the value of the damping factor ($D^A$) for the
$F_2$
slope  is
plotted. $D^A$ is defined in the following way:
\beq \label{D}
\frac{d F^A_2 (x,Q^2)}{d \ln Q^2}\,\,=\,\,D^A(x,Q^2)\,\,A\, \frac{d
F^{N;DGLAP}_2
(x,Q^2)}{d \ln Q^2}\,\,,
\eeq
where $F^{N;DGLAP}_2$ is the $F_2$ sructure function for a
nucleon in
the DGLAP approximation.
It turns out that the value of the effect is sizable for $Q^2 <
10\,GeV^2$ and strongly depends on $A$.


\section{Searching new observables.}

\subsection{Maxima in
$\mathbf{ \frac{ F_L}{F_T }}$ and  in
 $\mathbf{\frac{F^D_L}{F^D_T}}$.}
\begin{figure}
\begin{flushleft}
\begin{tabular}{cc}
$ $& $ $\\
\psfig{file=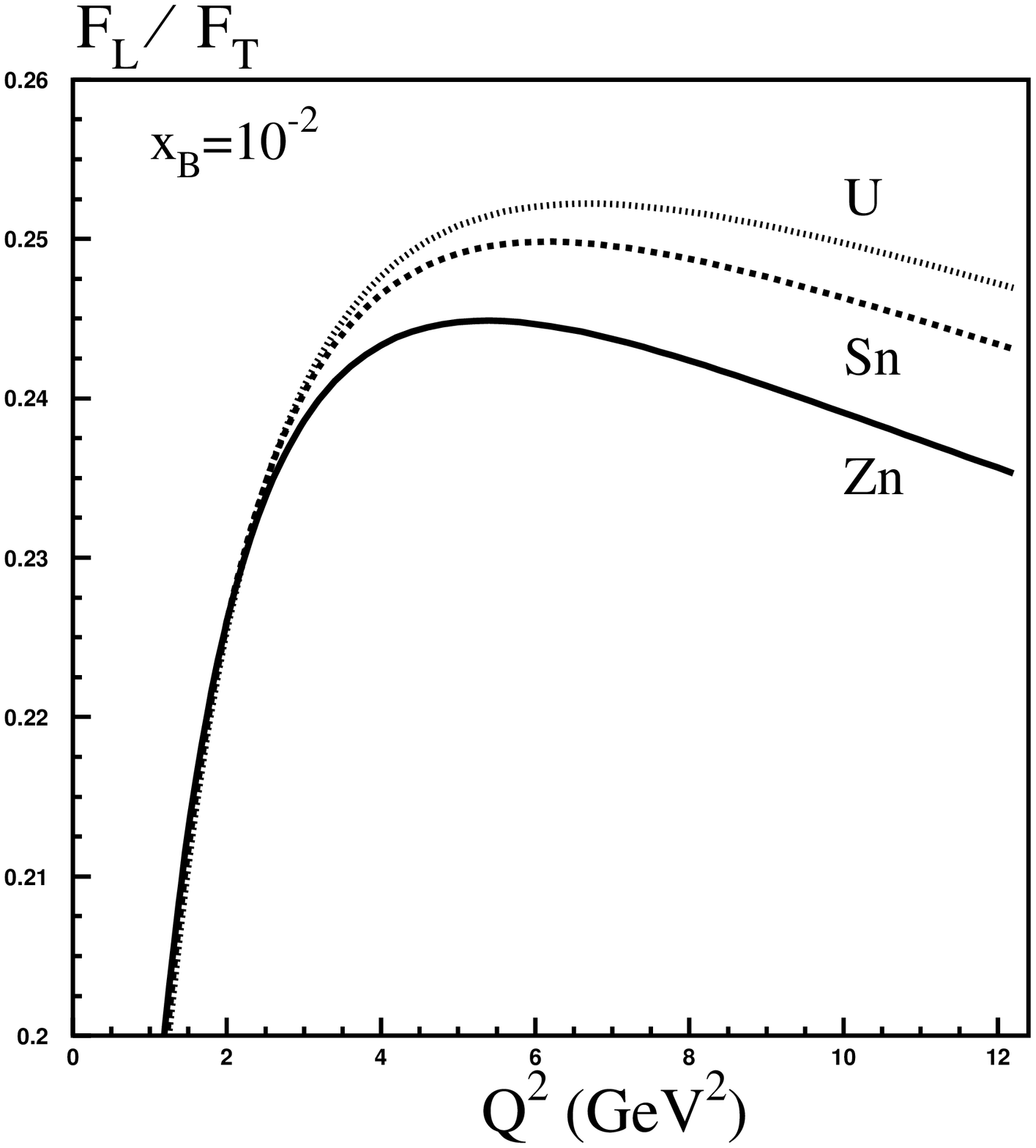,width=80mm,height=80mm}
&
\psfig{file=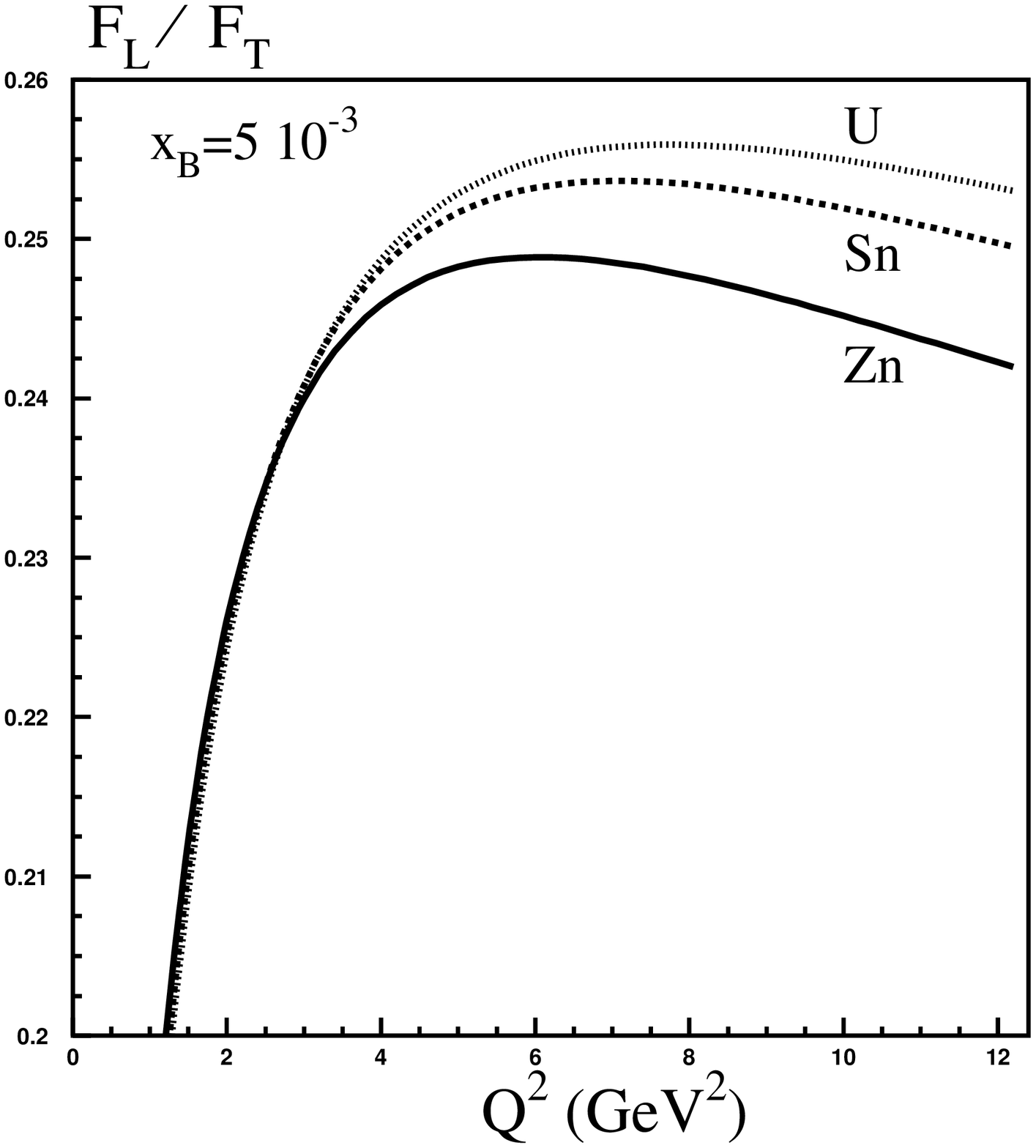,width=80mm,height=80mm}\\
$ $ & $ $\\
\psfig{file=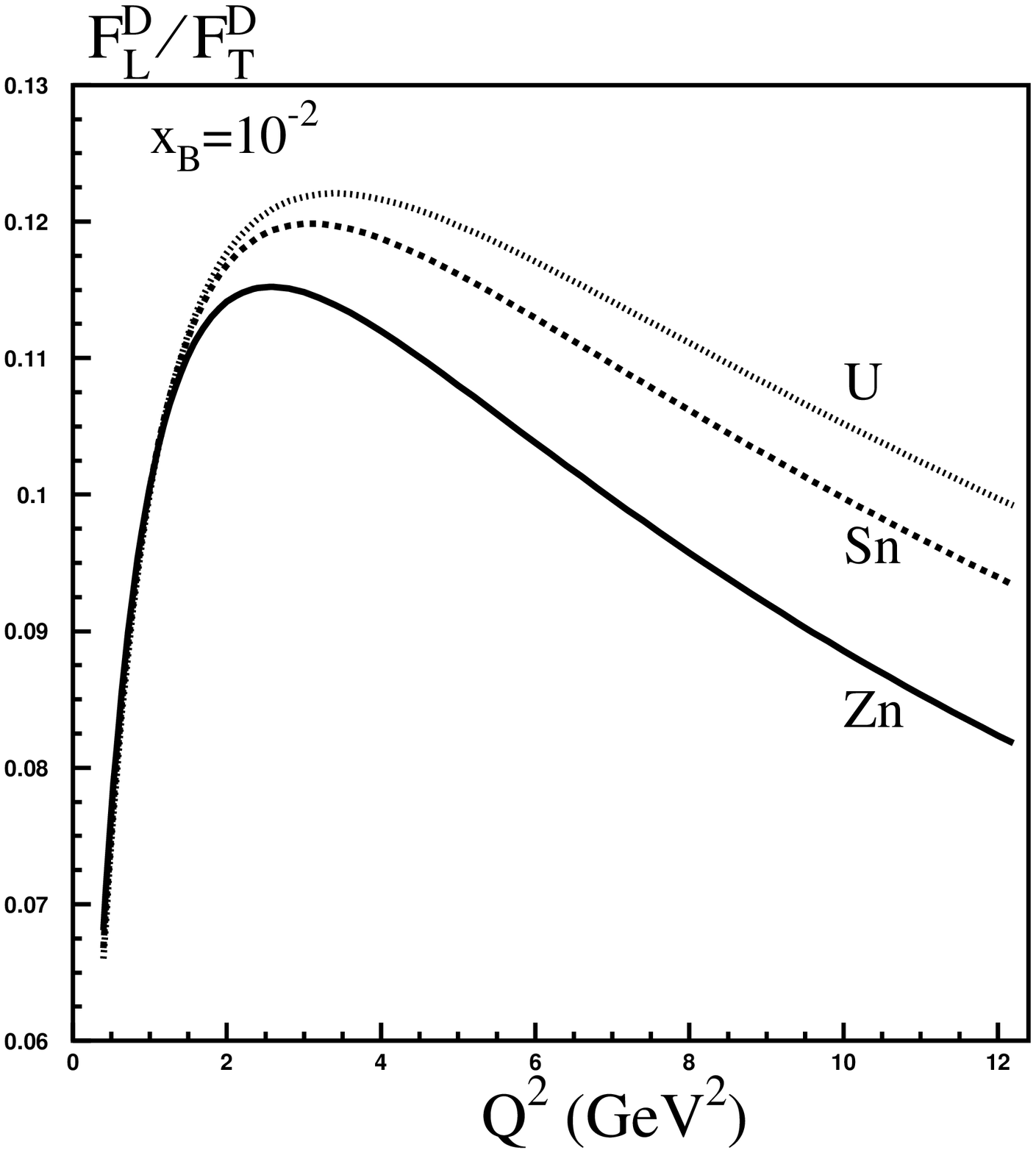,width=80mm,height=80mm}
&
\psfig{file=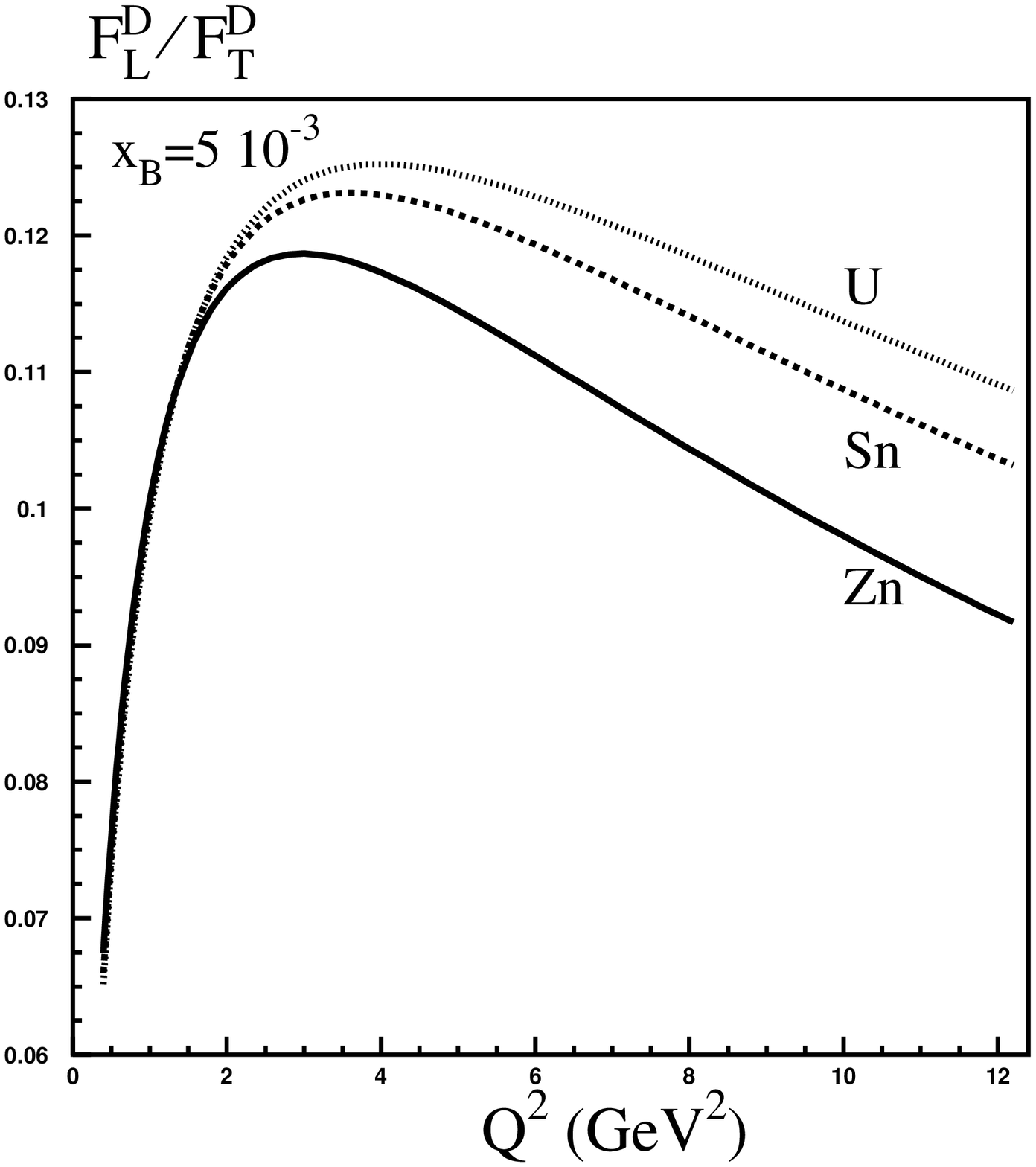,width=80mm,height=80mm}

\end{tabular}
\end{flushleft}
\caption{}
\label{max1}

\end{figure}

\begin{figure}
\begin{flushleft}
\begin{tabular}{cc}
\psfig{file=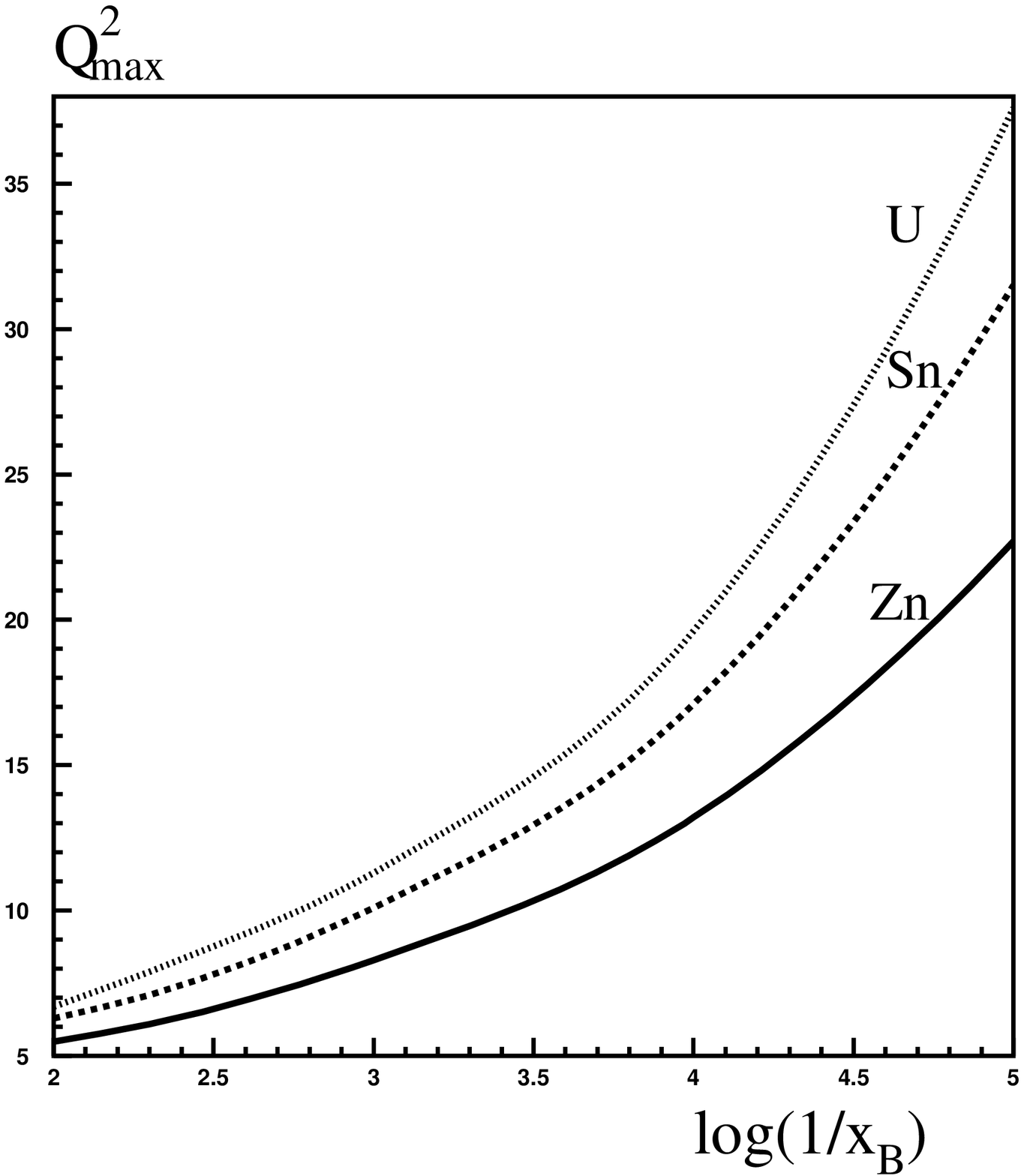,width=80mm,height=80mm}
&
\psfig{file=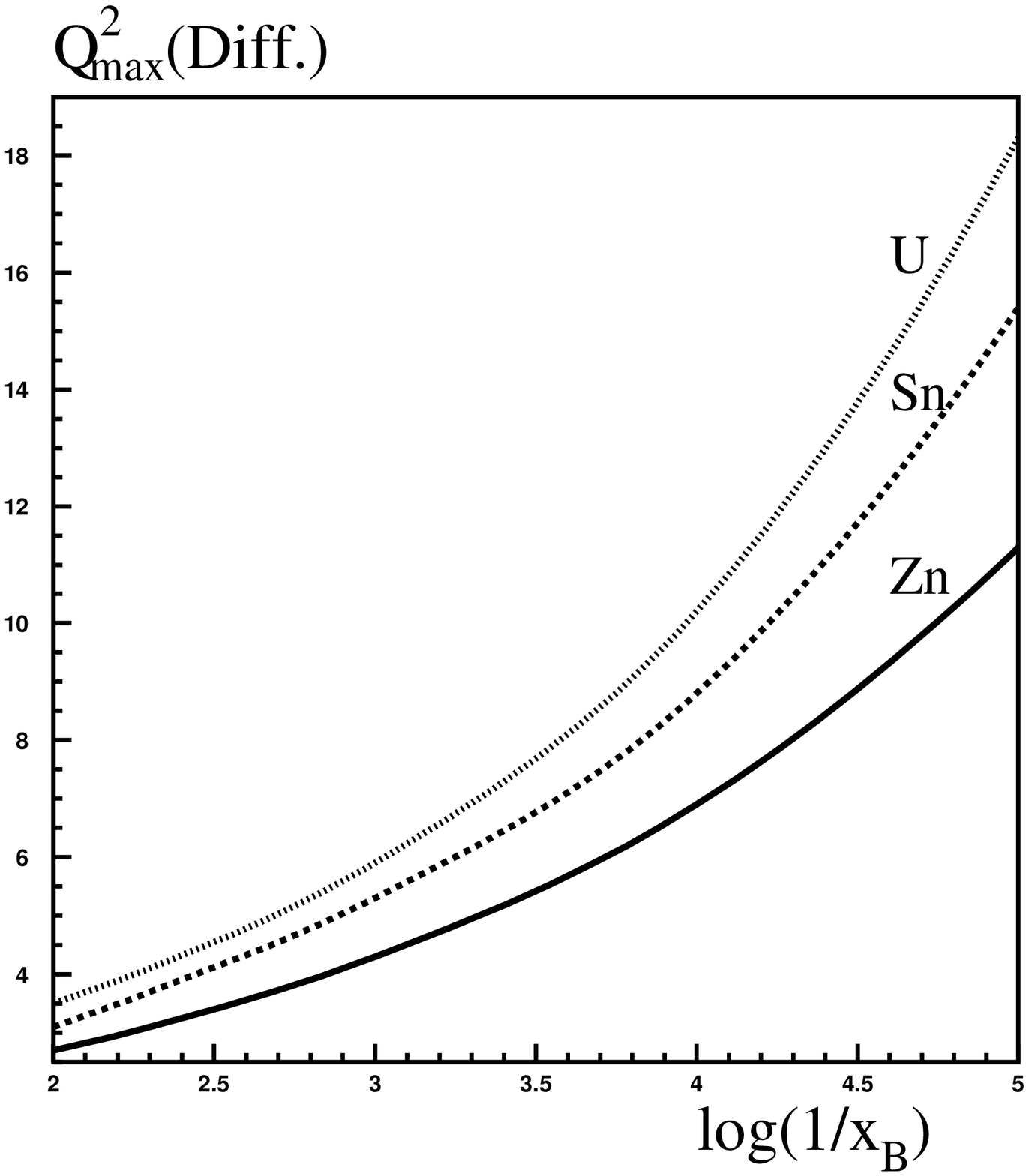,width=80mm,height=80mm}\\
\psfig{file=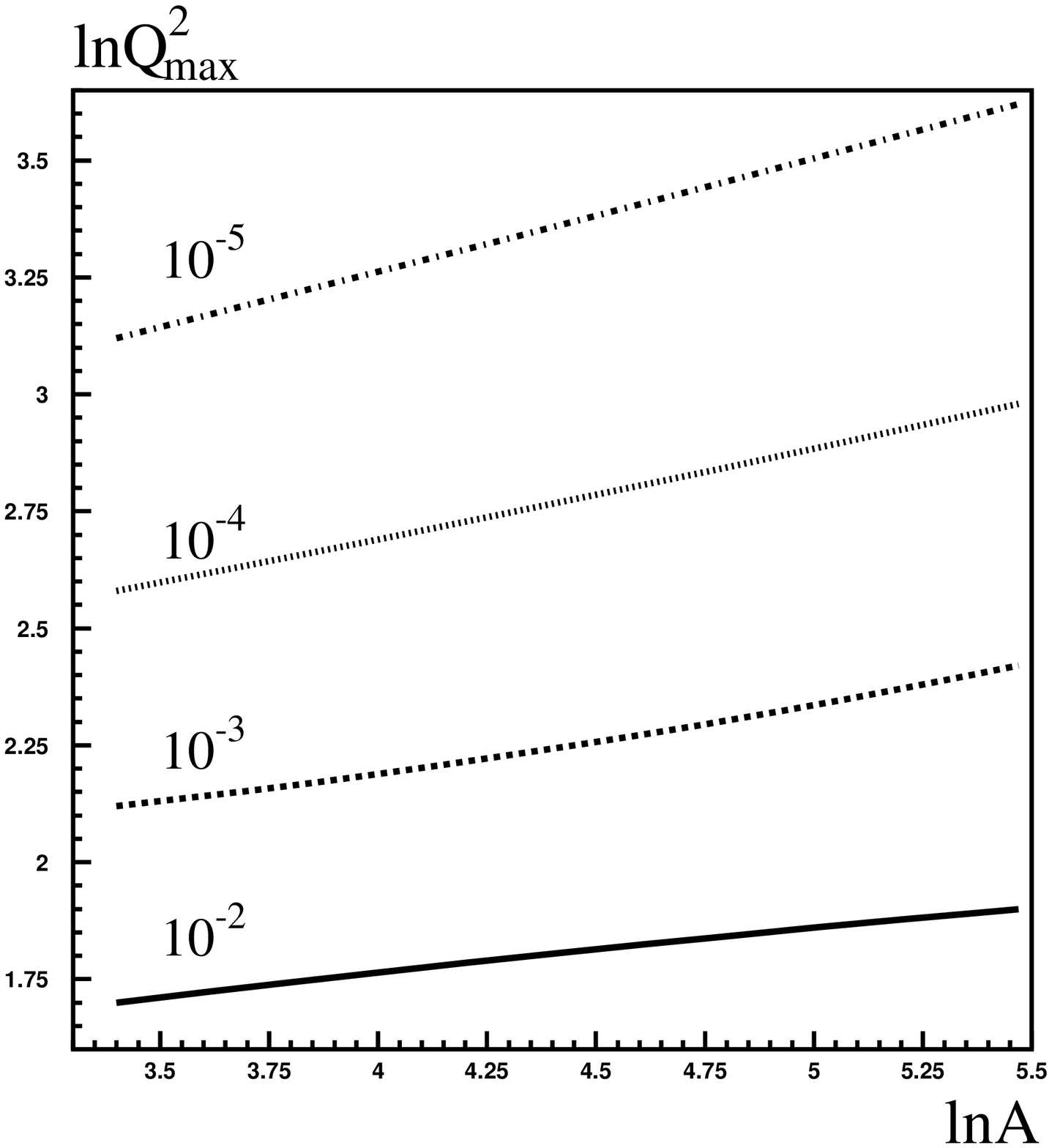,width=80mm,height=80mm}  
&
\psfig{file=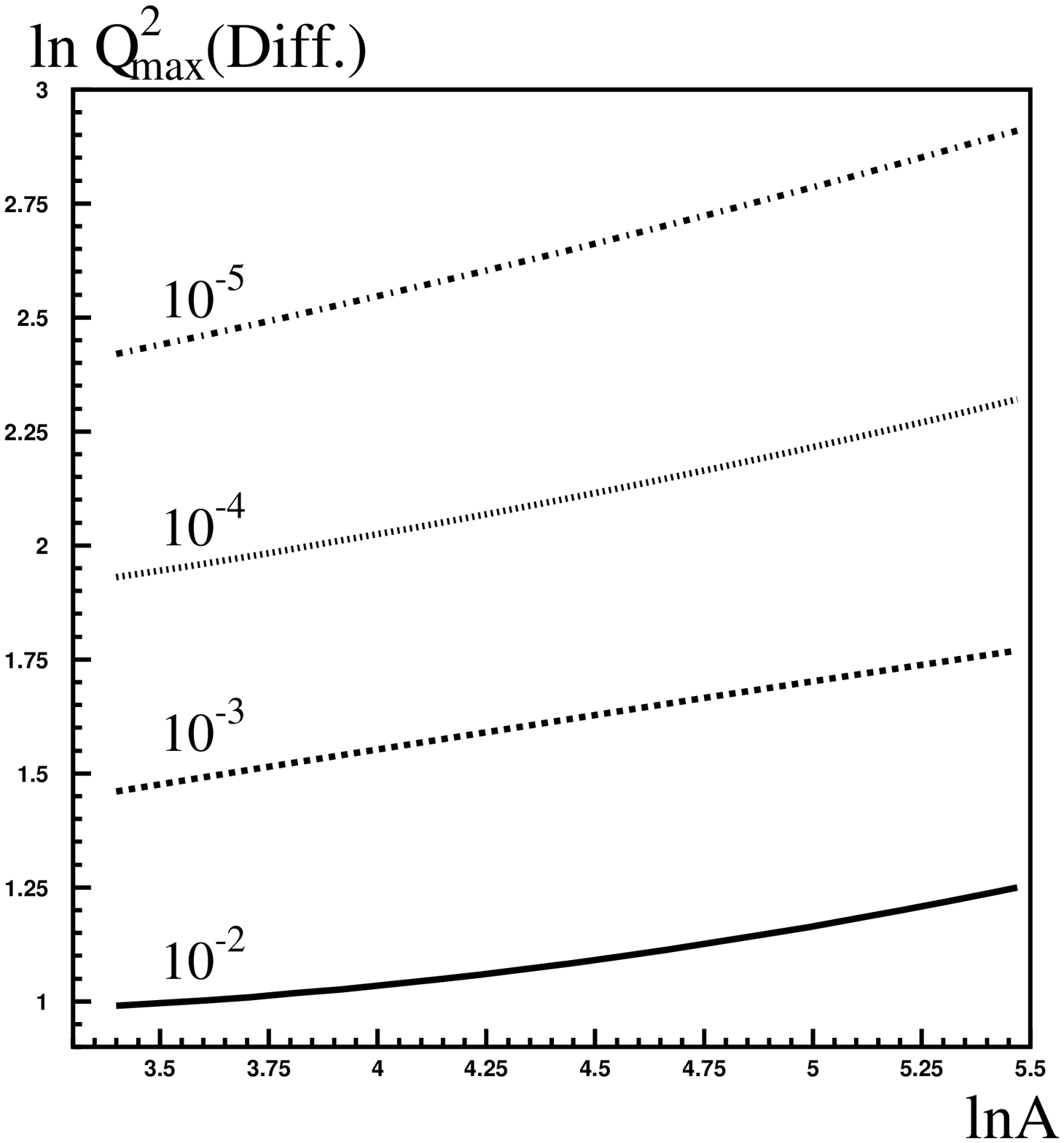,width=80mm,height=80mm}
\end{tabular}
\end{flushleft}
\caption{}
\label{max2}
\end{figure}

The main idea of this calculation\cite{MAXWE} is to show that these ratios
have
maxima at $Q^2 = Q^2_{max}(x;A)$, being plotted at fixed $x$. We want to
claim that $Q^2_{max}(x;A) \,\,\approx \,\,Q^2_s(x;A)$. Figs. \ref{max1}
and \ref{max2} show that such a suggestion can be right.

\subsection{Higher twists in $\mathbf{F_{L,T}}$ and in
$\mathbf{F^D_{L,T}}$.}

It is well known that any structure function ( $F_2(x,Q^2)$ for example )
can be written 
\beq \label{ST}
F_2(x,Q^2)\,\,=\,\,F^{LT}_2(x, ln Q^2 )\,\,+\,\,\frac{M^2}{Q^2}
\,F^{HT}_2(x, ln Q^2
)\,\,+\,\,...\,\,+\,\,\left(\,\frac{M^2}{Q^2}\,\right)^n\,F^{nT}_2(x, ln
Q^2 )\,\,+....
\eeq
Terms, which are small in terms  of $Q^2$ power , called  higher twist
contributions. The $ln Q^2$ dependence of the leading and higher twist
structure functions ( $F^{LT}_2$ and $ F^{HT}_2 $ in \eq{ST} ) is governed
by the evolution equations. The DGLAP evolution equations \cite{DGLAP}
give the $ln Q^2$ dependence of the leading twist structure function (
$F^{LT}_2$ ) only. Unfortunately, we know only a little about the  higher
twist
contributions.
\begin{enumerate}
\item \quad We know the evolution equations for all higher twist structure
functions \cite{LHT};
\item \quad We know the behaviour of the higher twist structure functions
at low $x$ \cite{HT}. For example,
$$
  F^{HT}_2 (x,ln Q^2 ) |_{x \ll 1} \,\,\longrightarrow\,\,F^{LT}_2(x,  ln
Q^2) \,\cdot\,xG^{LT}(x, ln Q^2 ) \,\,;
 $$
\item \quad We know, that higher twist contributions are needed to
describe the experimental data \cite{EXPHT}.
\end{enumerate}

However, it is difficult to estimate the value of the higher twist
contributions. Following Ref. \cite{BARTW}, we estimate the value of
different twist contribution for e A scattering. 

\begin{figure}
\begin{flushleft}
\begin{tabular}{ccc}
\multicolumn{3}{c}{\rule[-3mm]{0mm}{4mm} $ F_L(Q^2)\,\,A=238 (U)$}\\
$x_B=10^{-2}$&$x_B=5\cdot 10^{-3}$& $x_B=10^{-3}$\\[-10mm]
\psfig{file=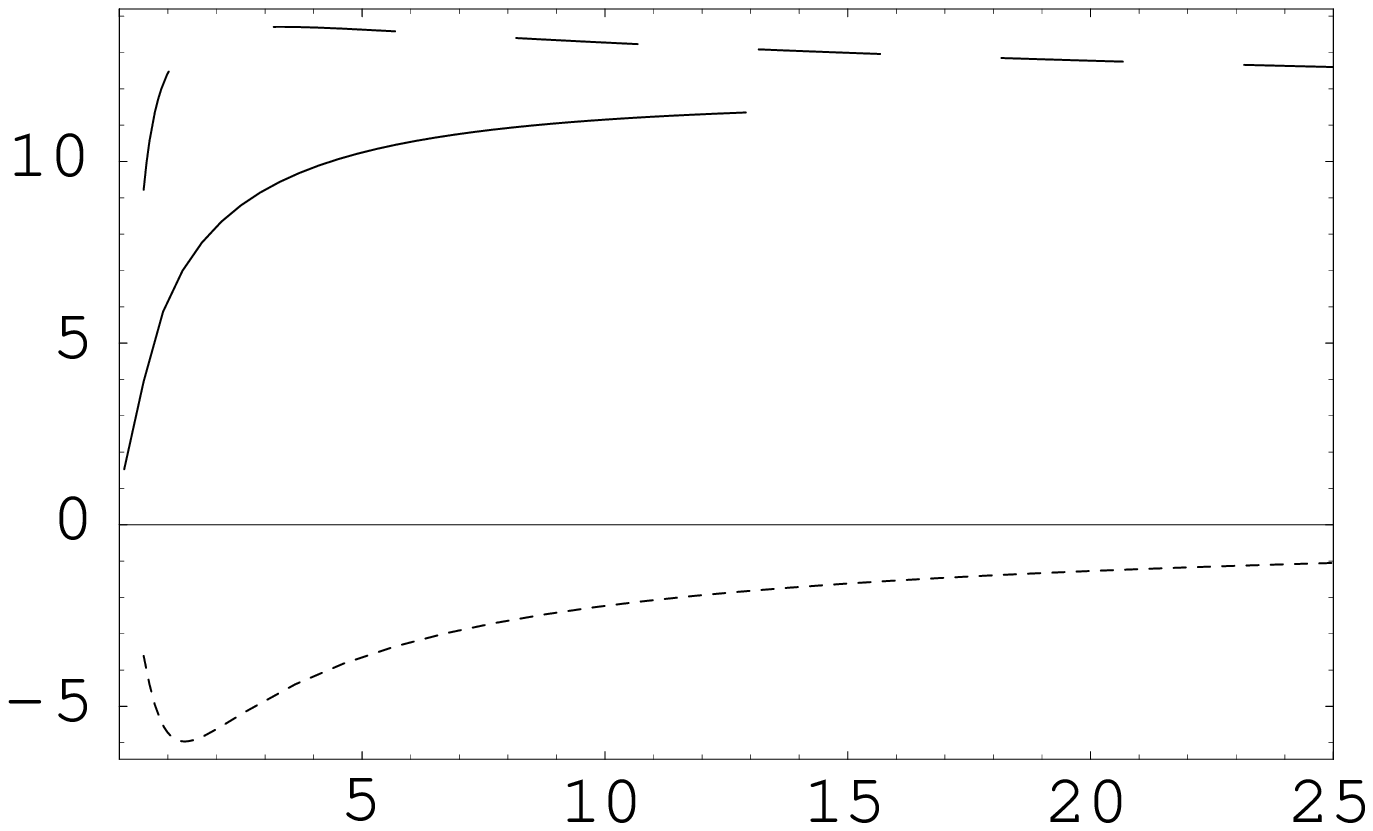,width=50mm,height=50mm} &
\psfig{file=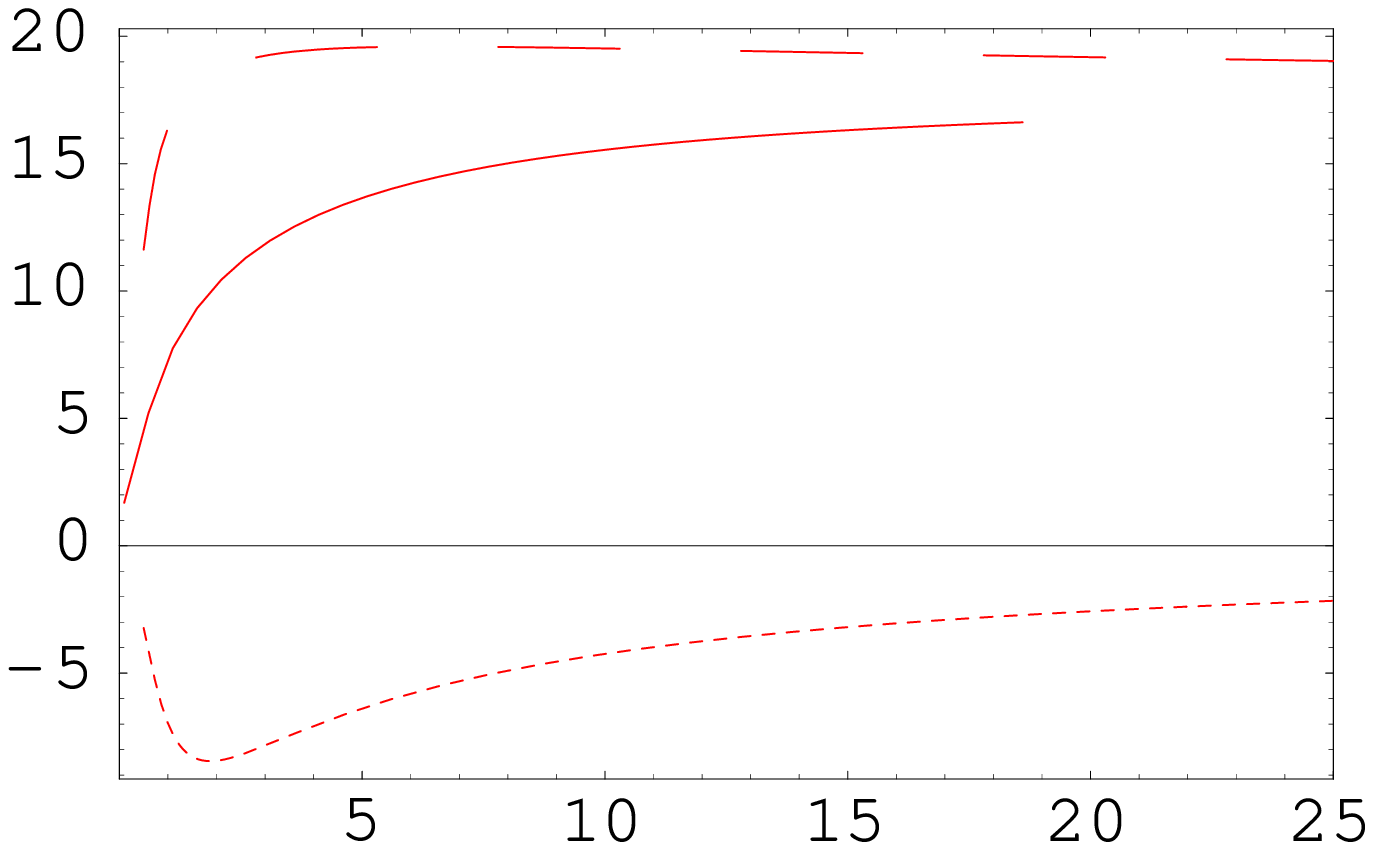,width=50mm,height=50mm}& 
\psfig{file=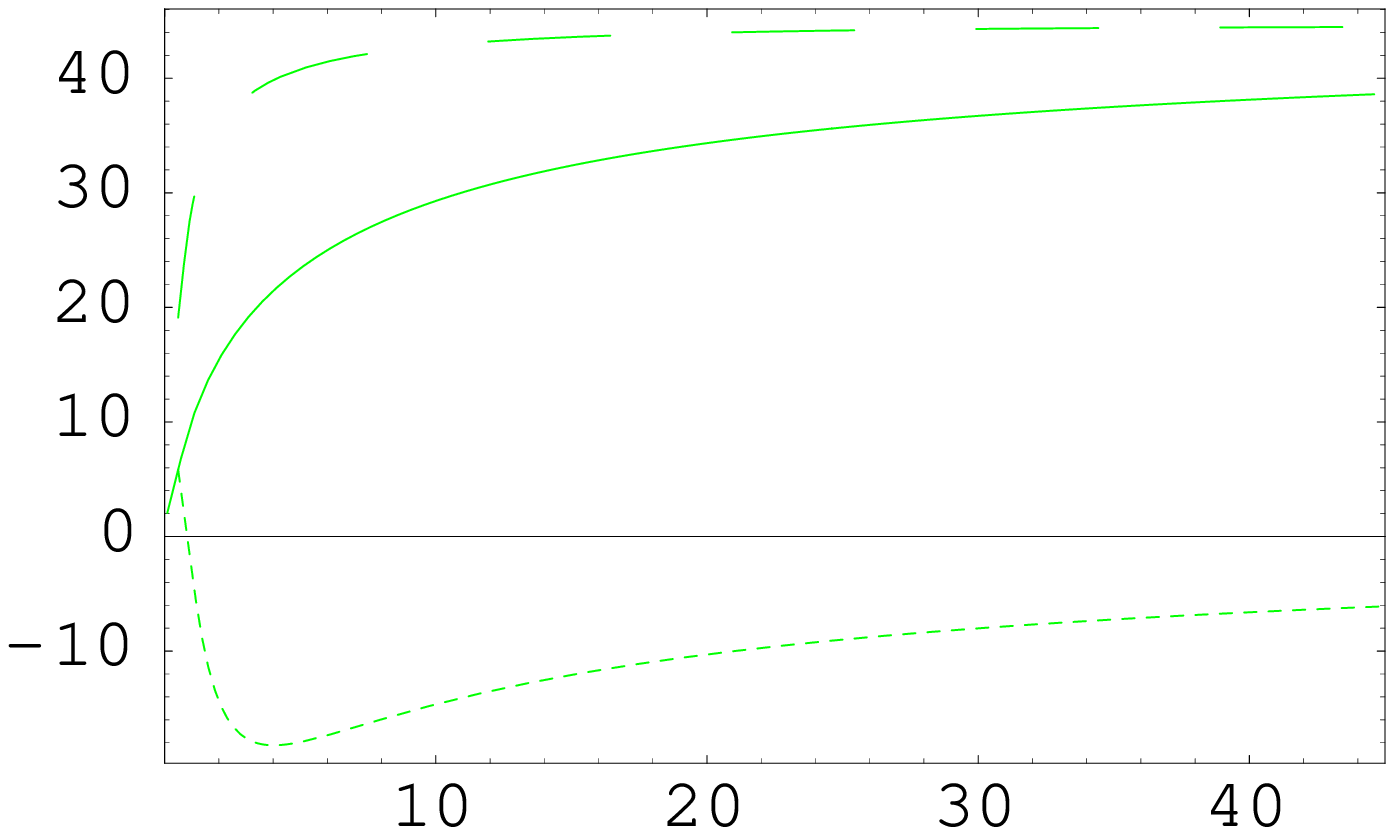,width=50mm,height=50mm}\\  
\multicolumn{3}{c}{\rule[-3mm]{0mm}{4mm} $ F_T(Q^2)\,\,A=238 (U)$}\\
$x_B=10^{-2}$&$x_B=5\cdot 10^{-3}$& $x_B=10^{-3}$\\[-10mm]
\psfig{file=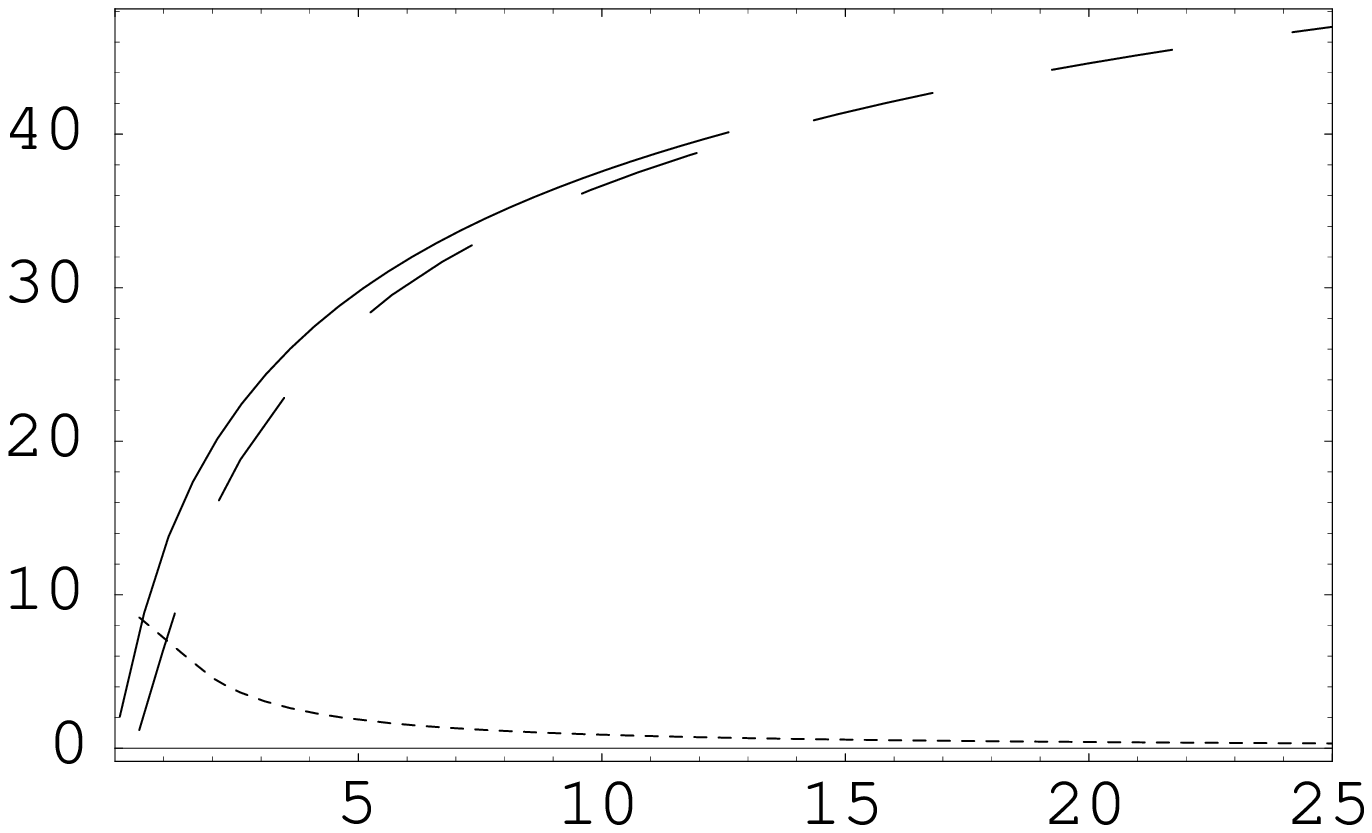,width=50mm,height=50mm} &
\psfig{file=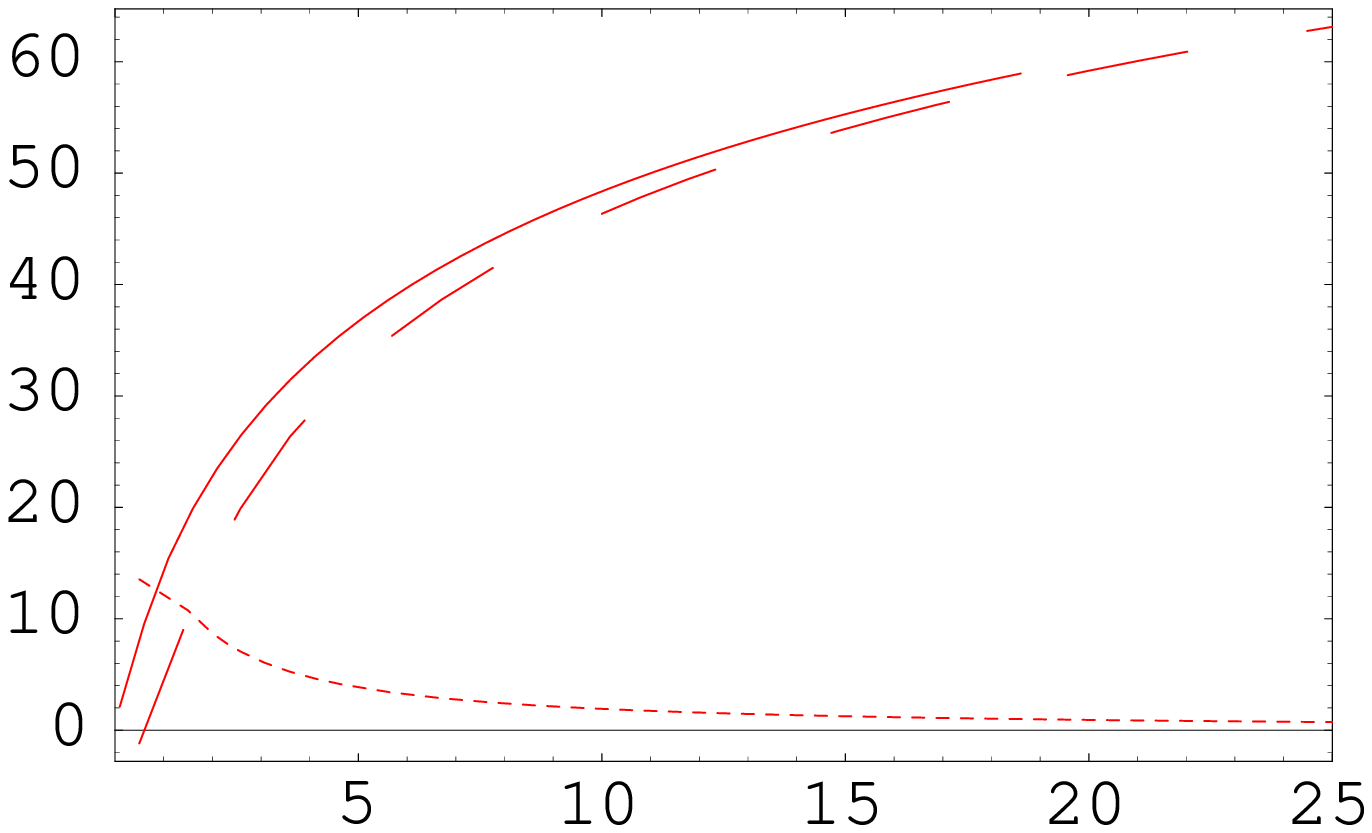,width=50mm,height=50mm} &
\psfig{file=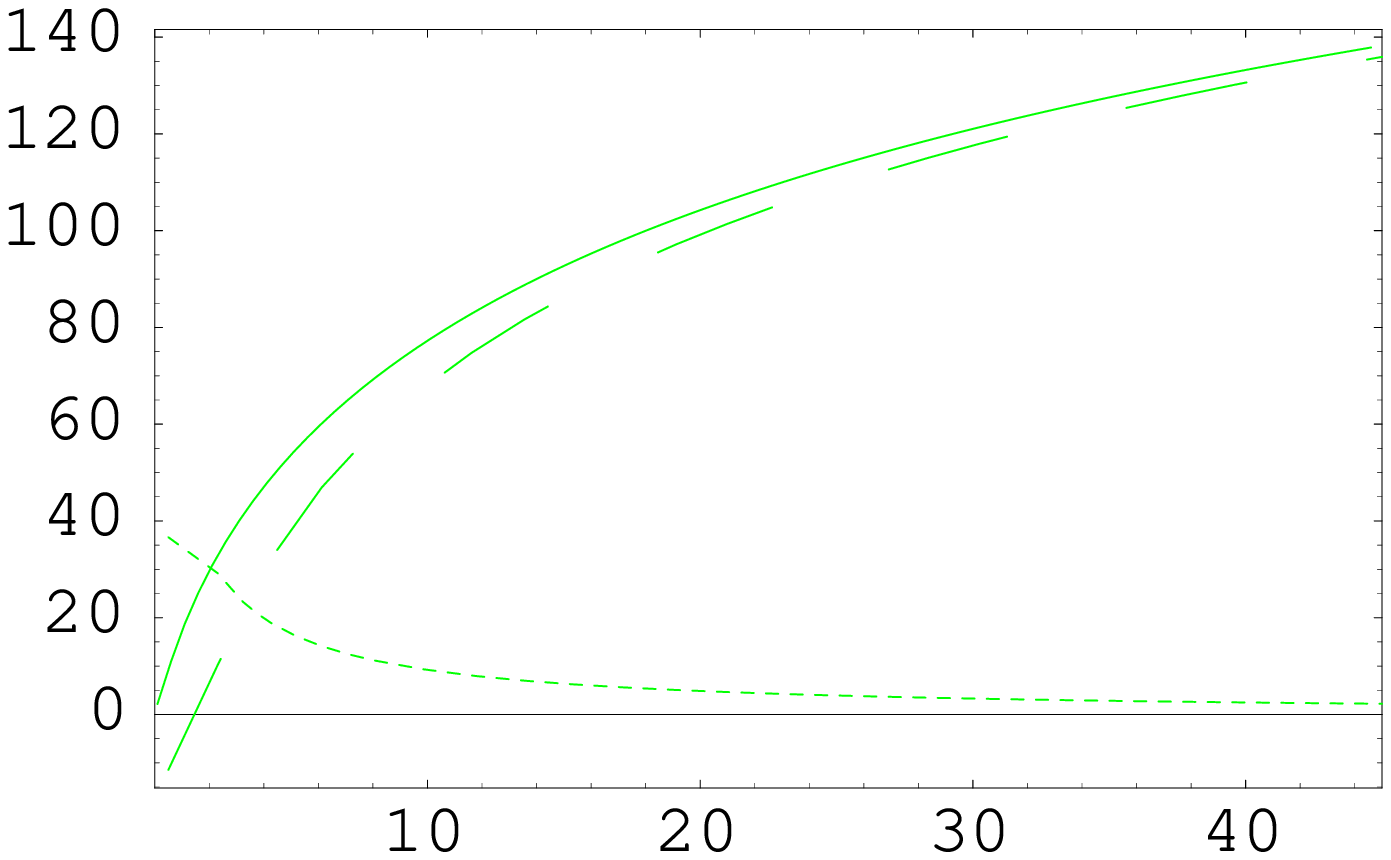,width=50mm,height=50mm}\\
\multicolumn{3}{c}{\rule[-3mm]{0mm}{4mm} $ F_L^D(Q^2)\,\,A=238 (U)$}\\
$x_B=10^{-2}$&$x_B=5\cdot 10^{-3}$& $x_B=10^{-3}$\\[-10mm]
\psfig{file=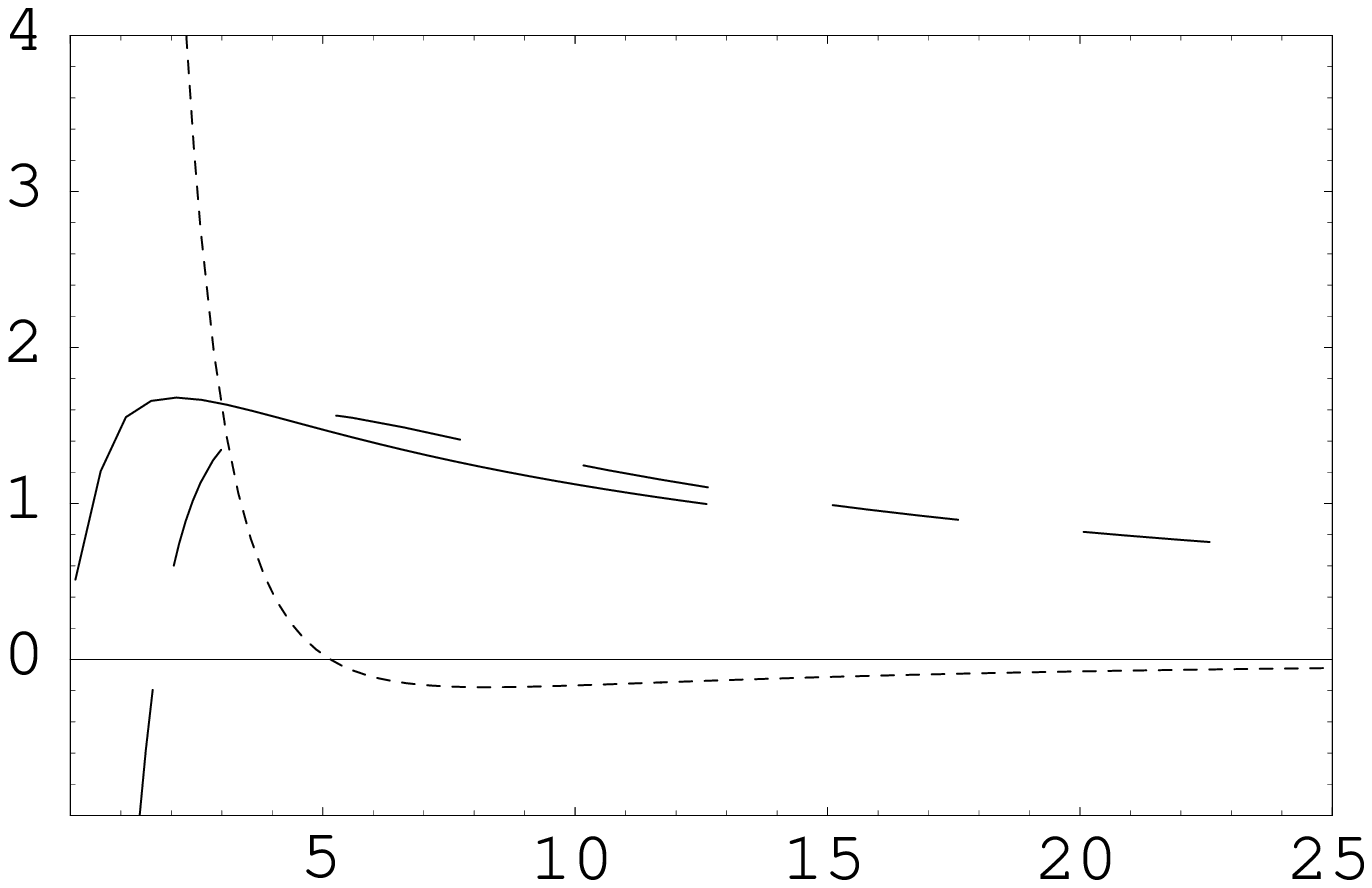,width=50mm,height=50mm} &
\psfig{file=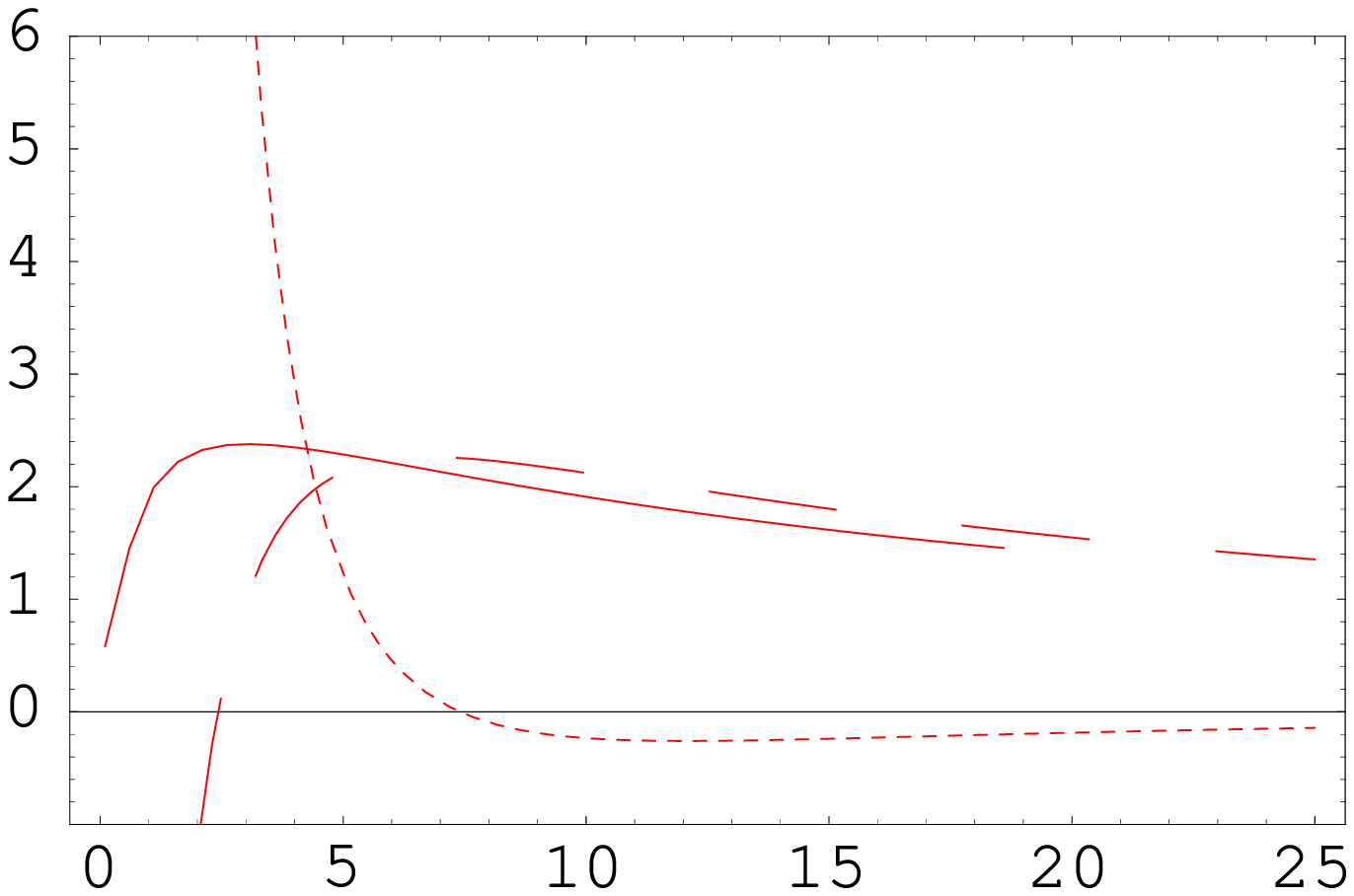,width=50mm,height=50mm} &
\psfig{file=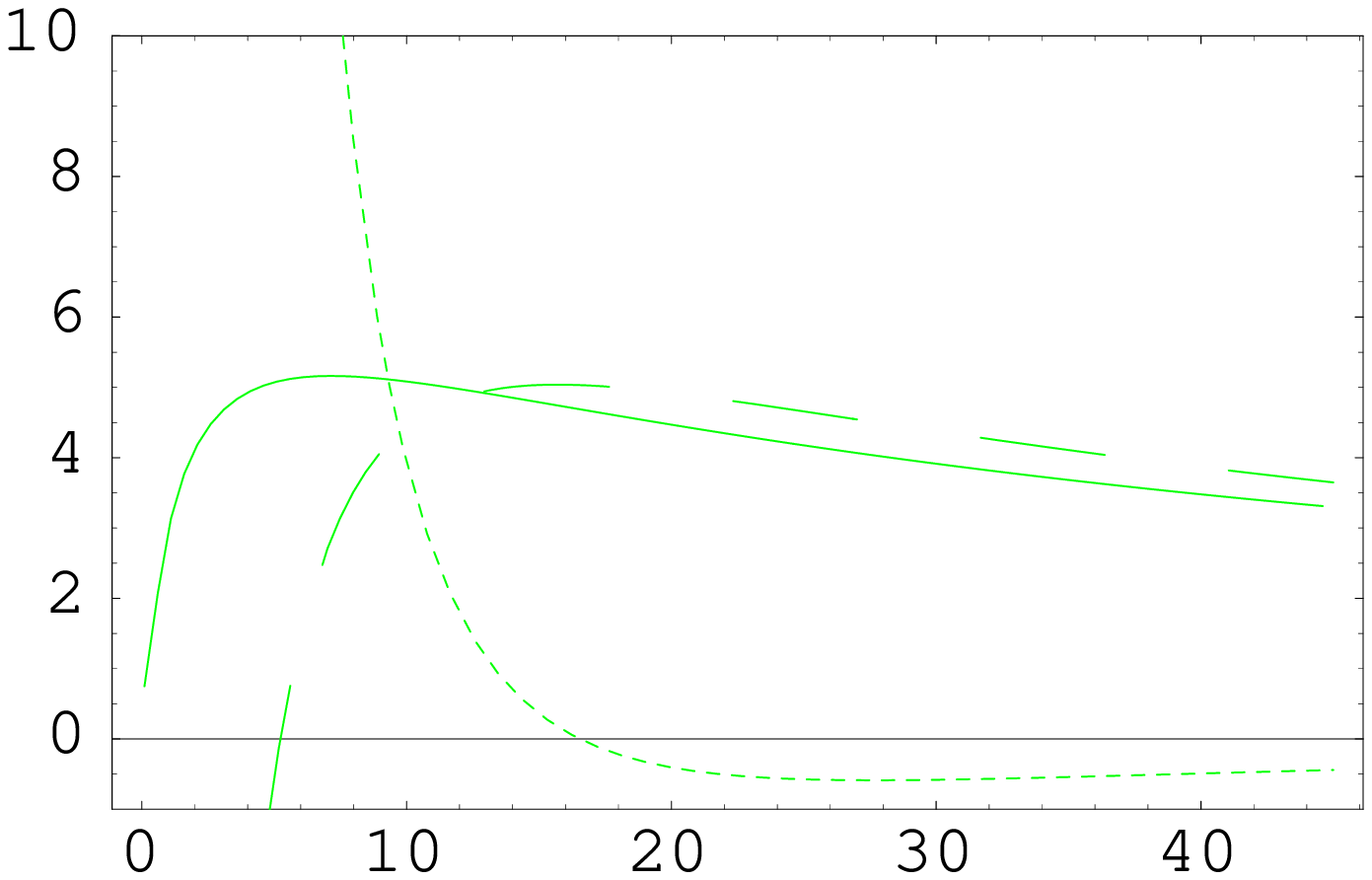,width=50mm,height=50mm}\\
\multicolumn{3}{c}{\rule[-3mm]{0mm}{4mm} $ F_T^D(Q^2)\,\,A=238 (U)$}\\
$x_B=10^{-2}$&$x_B=5\cdot 10^{-3}$& $x_B=10^{-3}$\\[-10mm]
\psfig{file=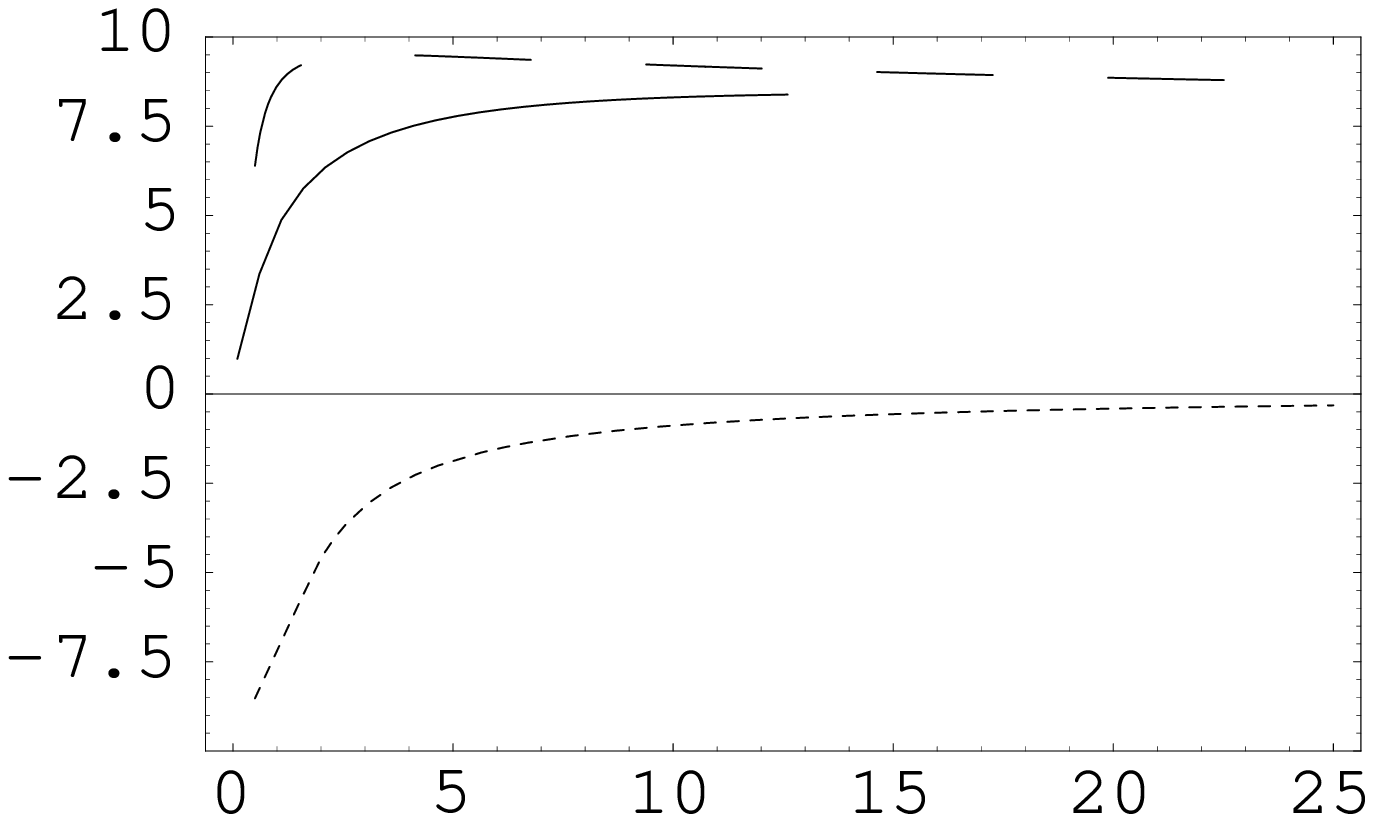,width=50mm,height=50mm} &
\psfig{file=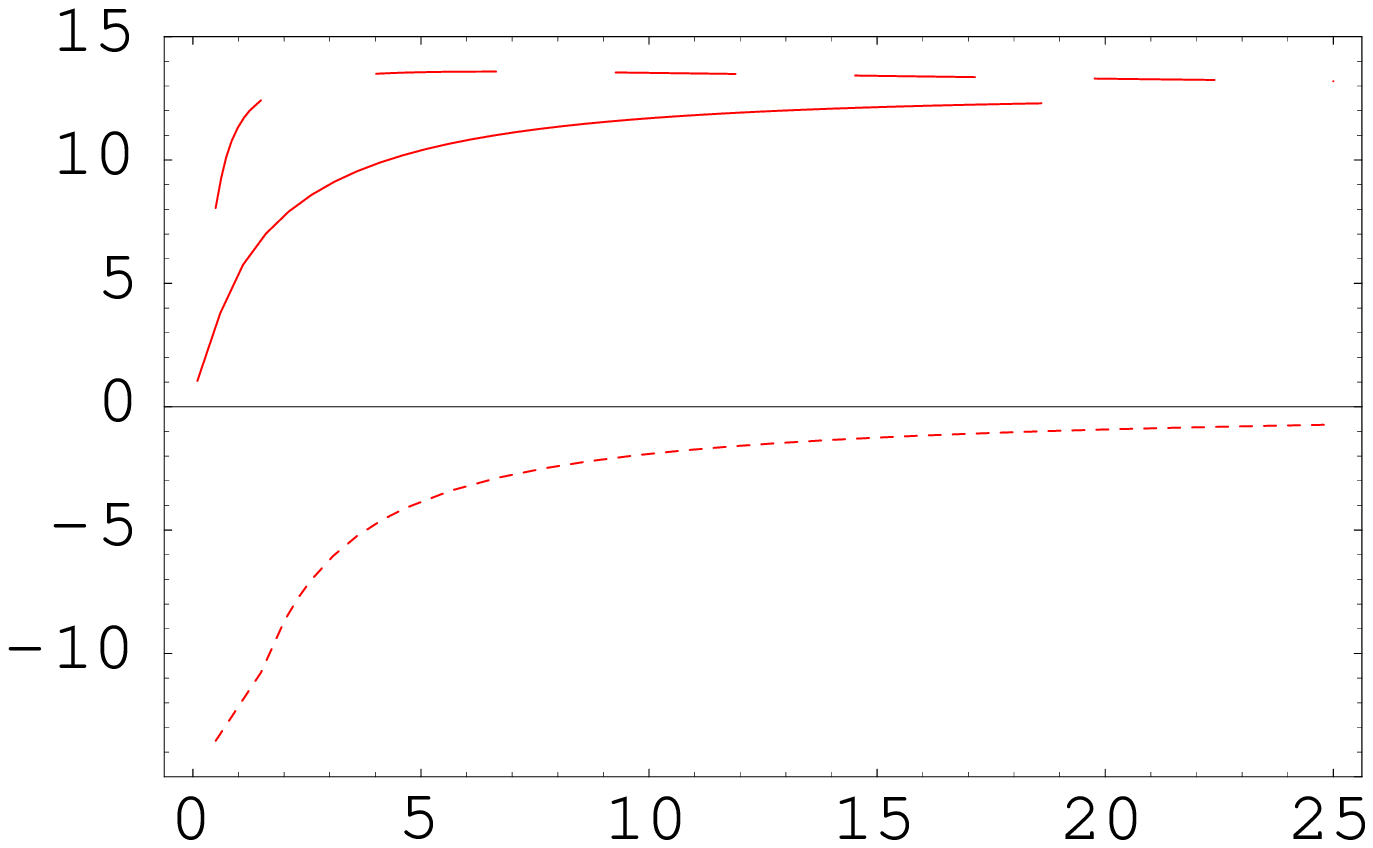,width=50mm,height=50mm} &
\psfig{file=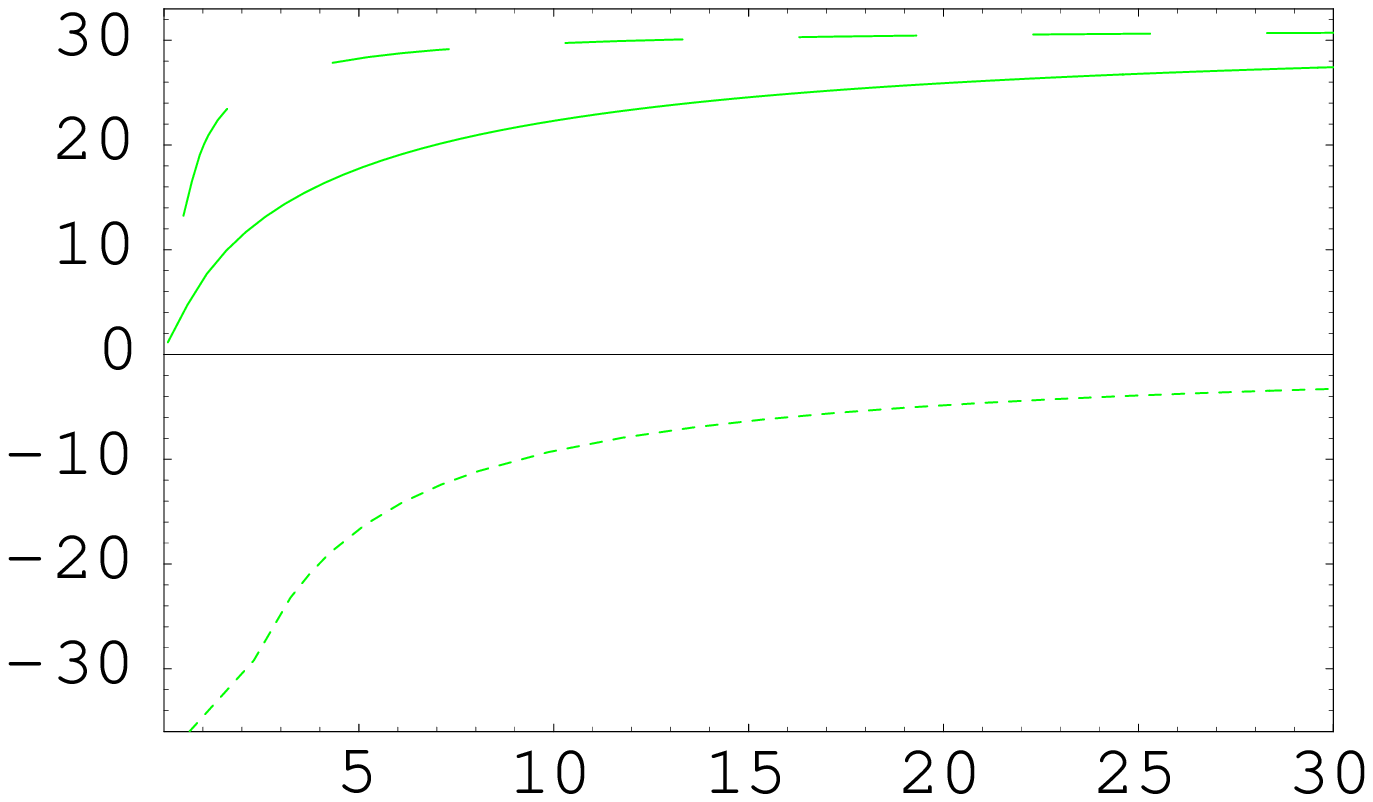,width=50mm,height=50mm}
\end{tabular}
\end{flushleft}
\caption{}
\label{tw}
\end{figure}

Fig.\ref{tw} shows that the higher twist contributions for nucleus target
become smaller
than the leading twist one only at $Q^2 > 5 GeV^2$ even at $x = 10^{-2}$.
It gives us a hope to treat them theoretically.
\section{Conclusions.}
\begin{enumerate}
\item\quad We have a solid theoretical approach for eA DIS,
 but  we need more experience in numerical solution of
the non-linear equation specifically for eRHIC kinematic region;
\item \quad  We know pretty well the scale of SC for eA interaction,
 but  we need more systematic study of DGLAP evolution for
nuclear structure functions and a special investigation whether the
initial parton distributions could be calculated for nuclear target from
the
initial parton distributions for  proton ;
\item \quad  Our estimates show that we will be able to see the
saturation scale in eA DIS at eRHIC being still far away from trivial
blackening of high energy  interaction with nuclei, but  we need
to check how close our model, which we use inn
practise, to theoretical estimates;

\item \quad  The $F^A_2$-slope is a very sensitive
observable for the
saturation scale,
 but , unfortunaly, we cannot expect a qualitatively different behaviou
for the saturation models in comparison with  others;

\item \quad   Maxima in ratios of $F_L/F_T$  and
$F^D_L/F^D_T$ give promising tool to extract the value of saturation scale
$Q_s(x;A)$, but  we need more study  on this subject and, in particular,
how the initial parton distribution for DGLAP evolution could affect our
predictions;

\item \quad  eA DIS is very instructive for separation of leading and
higher twist contributions, since the fact that typical momentum at which
these two contributions become of the same order  is   growing with $A$.
\end{enumerate}

\section*{Acknowledgments}

The authors are very much indebted to our coauthors Errol Gotsman, Larry
McLerran, Eran   
Naftali and Kirill Tuchin for their help and everyday discussions on the
subject.  E. L. thanks BNL Nuclear Theory group  and DESY  Theory group
for their hospitality and
creative atmosphere during several stages of this work.

 This research was supported in part by the BSF grant $\#$ 9800276 and by
Israeli Science Foundation, founded by the Israeli Academy of Science
and Humanities.

\begin{thebibliography}{99}
\bibitem{GRIB}
V.N. Gribov, {\it Sov. Phys. JETP} {\bf 30} (1970) 709.
\bibitem{GLMG}
E.  Gotsman,  E.  Levin and U.  Maor, {it Nucl. Phys. } {\bf B464} (1996)
251, {\bf B493} (1997) 354, {\bf B539} (1999) 535,
{\it Eur. Phys. J.} {\bf C5} (1998) 303, {\it Phys.Lett.} 
{\bf B425} (1998) 369;{\bf B403} (1997) 120;
E.  Gotsman,  E.  Levin, U.  Maor and E. Naftali,
{\it Nucl.Phys.} {\bf B539} (1999) 535, {\it Eur.Phys.J.}
{\bf C10} (1999) 689, {\bf C14} (2000) 511.

\bibitem{AGL}
A.L. Ayala Filho, M.B. Gay Ducati and E.  Levin, {\it Nucl. Phys.} {\bf
B493} (1997) 305, {\bf B511} (1998) 355, {\it Phys. Lett.} {\bf B388}
(1996) 188.

\bibitem{OSAKA}
E. Gotsman, E. Ferreira, E.
Levin, U. Maor and  E. Naftali, {\it `` Screening corrections in DIS at
low $Q^2$ and $x$"}, 
Talk given at 30th International Conference on High-Energy Physics
(ICHEP 2000), Osaka, Japan, 27 Jul - 2 Aug 2000; {\tt  hep-ph/0007274 }.  
\bibitem{MU94}
A.H.  Mueller, {\it  Nucl. Phys.}  {\bf B415} (1994) 373.
\bibitem{LF}
E. Levin and L. Frankfurt, {\it JETP Lett.} {\bf 2} (1965) 65;
H. J. Lipkin and F. Scheck, {\it Phys. Rev. Lett.} {\bf 16 } (1966) 71.
\bibitem{MU90}
A. H. Mueller, {\it Nucl. Phys.} {\bf B335} (1990) 115,
\bibitem{SAT}
L.V. Gribov, E.M.  Levin and M.G.  Ryskin, {\it Phys. Rep}
{\bf 100} (1983) 1;
A.H. Mueller and J. Qiu, {\it Nucl. Phys.} {\bf B268} (1986) 427;
L. McLerran and R. Venugopalan,{\it Phys. Rev. } {\bf D49} (1994)
2233,3352, {\bf 50} (1994) 2225, {\bf 53} (1996) 458, {\bf 59} (1999)
094002.
\bibitem{DGLAP}
V. N. Gribov and L. N. Lipatov, {\it Yad. Fiz} {\bf 15} (1972) 781;
L. N. Lipatov,{ Sov. Phys. J. Nucl. Phys.} {\bf 20} (1975) 94;
G. Altarelli and G. Parisi, {\it Nucl. Phys.} {\bf B126} (1977) 29;
Yu. L. Dokshitzer, {\it Sov. Phys. JETP} {\bf 46} (1977) 641.
\bibitem{MULA}
 J. Jalilian-Marian, A. Kovner, L. McLerran,
 and H. Weigert, { \it Phys. Rev.} {\bf  D  55}, (1997) 5414;
A.H. Mueller, {\it Phys.Lett.} {\bf B475} (2000)
220, {\it Nucl.Phys.} {\bf B572} (2000) 227.


\bibitem{KOLE}
Yu. V. Kovchegov,  { \it Phys.Rev.} {\bf D61} (2000) 074018;
 E. Levin and K. Tuchin, {\it Nucl.Phys.} {\bf B573} (2000) 833. 
 
\bibitem{PTHEORY}
E.   Levin and M.G. Ryskin, \prep{189}{267}{1990};
J.C.Collins and J. Kwiecinski, \npb{335}{90}{89};
J. Bartels, J. Blumlein and G. Shuler, \zpc{50}{91}{91};
E. Laenen and E. Levin, \arnps{44}{94}{199}
and references therein;
A.L. Ayala, M.B. Gay Ducati and E.M. Levin, \npb{493}{97}{305},
\npb{510}{98}{355};
Ia. Balitsky, {\it Nucl.Phys. } {\bf B463}  (1996) 99;
Yu. Kovchegov, \prd{54}{1996}{5463}, \prd{55}{1997}{5445},
\prd{60}{1000}{034008},{\it Phys. Rev.}
{\bf D61} (2000)074018; A.H. Mueller,
{\it Nucl. Phys.} {\bf B572}(2000)227,
\npb{558}{99}{285}; Yu. V. Kovchegov, A.H. Mueller,
\npb{529}{98}{451};  I.~Balitsky \npb{463}{96}{99}; E.  Levin
and
K. Tuchin, {\it Nucl. Phys.} {\bf B573}(2000) 833;
\bibitem{ELTHEORY}
J. Jalilian-Marian, A. Kovner, L. McLerran  and  H.
Weigert, \prd{D55}{97}{5414};
J. Jalilian-Marian, A. Kovner and  H.
Weigert, \prd{59}{99}{014015};
J. Jalilian-Marian, A. Kovner and  H.
Weigert, \prd{59}{99}{014015};
J. Jalilian-Marian, A. Kovner, A.
Leonidov and  H. Weigert, \prd{59}{99}{014014,034007},
Erratum-ibid. \prd{59}{99}{099903};
A. Kovner, J.Guilherme Milhano and  H. Weigert,
OUTP-00-10P,NORDITA-2000-14-HE, {\tt hep-ph/0004014};
 H. Weigert, NORDITA-2000-34-HE, {\tt hep-ph/0004044};
M.~Braun LU-TP-00-06,{\tt hep-ph/0001268}.

\bibitem{EQ}
Ia. Balitsky, {\it Nucl.Phys. } {\bf B463}  (1996) 99;
Yu. Kovchegov,   
\prd{60}{1000}{034008}.
\bibitem{HERADATA}
A. M. Cooper-Sarkar, R. C. E. Devenish and A. De Roeck, {\it Int. J. Mod.
Phys.} {\bf A13} (1998) 33; H. Abramowicz and A. Caldwell, {\it Rev. Mod.
Phys.} {\bf 71} (1999) 1275.
\bibitem{DL} A. Donnachie, P. V.  Landshoff,{\it  Phys. Lett.} {\bf
B296} (1992)
  227;  {\bf B437}(1998)  408  and references therein.
\bibitem{GW}
K. Golec-Biernat and M. Wusthoff, {\it Phys. Rev.} {\bf D59} (1999)
014017; {\bf D60} (1999) 114023;
K.Golec-Biernat,{Talk 
at 8th International Workshop on Deep Inelastic Scattering and
QCD (DIS 2000)}, Liverpool, England, 25-30 Apr 2000,{\tt  hep-ph/0006080}. 
\bibitem{MK}
Yu. V. Kovchegov and L. McLerran, {\it Phys.Rev.} {\bf D60} (1999) 054025.
\bibitem{DDOUR}
E. Gotsman, E. Levin, M.
Lublinsky, U. Maor and  K. Tuchin, TAUP-2605-99,
{\tt  hep-ph/9911270}.
\bibitem{MAXWE}
E. Gotsman, E. Levin, U.
Maor, L. McLerran and K. Tuchin,
TAUP-2638-200, BNL-NT-00-19, Jul 2000,{\tt hep-ph/0007258}.
\bibitem{LHT}
A.P. Bukhvostov,G.V. Frolov, L.N. Lipatov and E.A. Kuraev, {\it Nucl.
Phys.} {\bf B258} (1985) 601.
\bibitem{HT}
J. Bartels, {\it Phys. Lett.} {\bf B298} (1993) 204, {\it Z. Phys.} {\bf
C60} (1993) 471; E.M. Levin, M.G. Ryskin and A.G. Shuvaev, {\it Nucl.
Phys.} {\bf B387} (1992) 589.
\bibitem{EXPHT}
 U.K. Yang and  A. Bodek,{\it  Phys.Rev.Lett.}
 {\bf 84} (2000) 5456, {\bf 82}
(1999) 2467 and references thertein.
\bibitem{BARTW}
J.Bartels, K. Golec-Biernat and  K. Peters,  DESY-00-038, Mar
2000,{\tt  hep-ph/0003042 }.


 
 
\end{thebibliography}
\end{document}